\documentclass[11pt]{article}

\usepackage[margin=1in]{geometry}
\usepackage{amssymb}
\usepackage{amsmath}
\usepackage{booktabs}
\usepackage{multirow}
\usepackage{rotating}
\usepackage{graphicx}
\usepackage[round,authoryear]{natbib}
\usepackage[hyphens]{url}
\usepackage{hyperref}

\hypersetup{
  colorlinks=true,
  linkcolor=blue,
  citecolor=blue,
  urlcolor=blue,
  pdftitle={Decomposing Crowd Wisdom: Domain-Specific Calibration Dynamics in Prediction Markets},
  pdfauthor={Nam Anh Le}
}
\begin{document}

\title{Decomposing Crowd Wisdom: Domain-Specific Calibration Dynamics in Prediction Markets}

\author{Nam Anh Le\thanks{Corresponding author: \href{mailto:me@namanhle.com}{me@namanhle.com}.}\\
\small National Economics University, Vietnam\\[0.5em]
\small Code and replication materials:
\href{https://github.com/namanhzz/prediction-market-calibration}{github.com/namanhzz/prediction-market-calibration}}
\date{}

\maketitle

\begin{abstract}
Prediction market prices are often read as probabilities, but this reading requires calibration. Using 353 million trades across 429,000 binary contracts on Kalshi and Polymarket, this paper measures how calibration varies with event domain, time-to-resolution and trade size. A descriptive decomposition of cell-level logistic recalibration slopes explains 87.3\% of in-sample variance on Kalshi (71.5\% out-of-sample). The most robust pattern is persistent underconfidence in political markets, where prices compress toward 50\%; it replicates on Polymarket. Large political trades on Kalshi are associated with further compression---a calibration-slope gap of roughly one-half that survives market- and event-clustered bootstraps but is not robust on Polymarket. A Bayesian measurement-error model that propagates first-stage uncertainty agrees with these conclusions and indicates that---under conservative event-clustered standard errors---roughly half of the raw slope variation reflects estimation noise. Calibration is therefore conditional: a price's meaning depends on what, when and how much is traded.
\end{abstract}


\vspace{1em}

%% ============================================================
%% SECTION 1: INTRODUCTION
%% ============================================================
\section{Introduction}\label{sec:intro}

On the evening of 5 November 2024, prediction market users watched US presidential election contracts trade near 62 cents. News coverage often translates such prices directly into probabilities. The translation is natural but not automatic: if a market is well \emph{calibrated}, contracts that trade at 62 cents should resolve affirmatively about 62\% of the time. If calibration varies across market settings, the same price can carry different probabilistic content in different contexts.

This paper asks how prediction market calibration varies with \emph{what} is being predicted, \emph{when} the price is observed and \emph{how much} is traded. The empirical object is not simply whether prediction markets display a favorite--longshot bias in the aggregate. Recent work already documents related biases and microstructure patterns on Kalshi and Polymarket \citep{becker2026microstructure,burgi2025makers,reichenbach2025polymarket}. The contribution here is to measure calibration as a structured function of domain, time-to-resolution and trade size, and to compare which patterns are stable across exchanges and which appear platform-specific.

Using 353 million trades across 429,000 binary contracts, the analysis treats Kalshi as the primary data source and Polymarket as a cross-platform comparison. The results are descriptive rather than causal. They show that political markets are persistently compressed toward 50\%, that short-horizon weather markets move in the opposite direction, and that Kalshi's large political trades are associated with additional compression that is not estimated as precisely on Polymarket. These patterns matter for users of prediction market prices because calibration is conditional: a single price is more informative when its historical domain, horizon and trading context are known.

The paper proceeds as follows. Section~\ref{sec:related} positions the analysis relative to prediction market efficiency, calibration and microstructure research. Section~\ref{sec:data} describes the exchanges, data and definitions. Section~\ref{sec:measuring} introduces the calibration estimators and empirical design. Section~\ref{sec:landscape} presents the descriptive calibration landscape. Sections~\ref{sec:artefacts}--\ref{sec:bayesian} decompose the patterns and assess robustness. Section~\ref{sec:discussion} discusses interpretation and limitations, and Section~\ref{sec:conclusion} concludes.

\section{Related work}\label{sec:related}

Prediction markets are financial exchanges where contracts pay \$1 if a specified event occurs and \$0 otherwise \citep{forsythe1992anatomy,rhode2004historical}. A contract trading at price $p \in (0,1)$ is commonly interpreted as encoding a crowd-sourced probability $p$ that the event will happen. This interpretation rests on theoretical work linking equilibrium prices to aggregated beliefs \citep{wolfers2004prediction,manski2006interpreting,arrow2008promise}, while \citet{manski2006interpreting} shows that prices need not equal mean beliefs under heterogeneous preferences. The broader case for decentralized information aggregation draws on \citet{hayek1945use}, \citet{galton1907vox}, \citet{surowiecki2004wisdom} and related work on diverse groups \citep{hong2004groups}. Prediction markets have been studied in elections \citep{forsythe1992anatomy,berg2008prediction,rothschild2009forecasting}, organizations \citep{cowgill2015corporate}, public policy and scientific replication \citep{dreber2015using,camerer2016evaluating}.

The calibration literature provides the measurement framework. Calibration means that events assigned probability $p$ occur with empirical frequency $p$ \citep{brier1950verification,murphy1977reliability,lichtenstein1982calibration}. Logistic recalibration \citep{platt1999probabilistic,niculescumizil2005predicting,gneiting2007strictly} summarizes deviations through an intercept and slope in logit space. Related work studies probability aggregation and recalibration using logit models \citep{satopaa2014combining,guthrie2024boldness,palley2024robust}. In betting and prediction markets, the favorite--longshot bias is a central empirical regularity \citep{griffith1949odds,ali1977probability,thaler1988anomalies,snowberg2010explaining,page2013prediction}. \citet{ottaviani2008favorite} emphasize that there are multiple explanations for this bias, including risk preferences, probability misperception, information asymmetries and market mechanisms; no single explanation should be assumed without market-specific evidence.

The closest recent papers are \citet{becker2026microstructure}, \citet{burgi2025makers} and \citet{reichenbach2025polymarket}. These studies show that modern event-contract exchanges are informative but exhibit economically meaningful biases and trader heterogeneity. This paper differs by asking where calibration variation is concentrated across domain, horizon and trade size, and by comparing Kalshi and Polymarket within a common recalibration framework. The domain-specific focus is therefore not a claim that earlier work ignored all heterogeneity; many prior prediction-market and forecasting studies are intentionally single-domain. The claim is narrower: pooling domains can hide large differences in calibration dynamics.

Market microstructure research helps interpret the trade-size results. In standard models, larger trades can convey private information and improve calibration \citep{kyle1985continuous,glosten1985bid,easley1987price}. The empirical finding here goes in the opposite direction for Kalshi political markets: larger trades are associated with more compressed prices. That pattern is treated as a diagnostic fact requiring institutional explanation, not as direct evidence of a particular causal mechanism.

The forecasting tournament literature provides an additional benchmark. The Good Judgment Project \citep{mellers2014psychological,tetlock2015superforecasting} demonstrates that structured forecasting teams can achieve strong calibration, and \citet{baron2014two} show that extremizing transformations can improve underconfident forecasts. More recent platforms and datasets \citep{karger2023reciprocal} similarly show that calibration depends on the structure of the forecasting problem. The present paper brings that conditional perspective to transaction-level prediction market prices.
%% ============================================================
%% SECTION 2: DATA
%% ============================================================
\section{Market mechanics, data and definitions}\label{sec:data}

\subsection{Data source}\label{sec:datasource}

Data are drawn from two major prediction market exchanges using the data collection framework of \citet{becker2026microstructure}. The primary data source is Kalshi, regulated by the US Commodity Futures Trading Commission (CFTC), which operates a central limit order book for binary event contracts paying \$1.00 if a specified event occurs and \$0.00 otherwise. Contracts trade at prices between \$0.01 and \$0.99. For cross-platform comparison, the analysis uses Polymarket, a decentralized exchange operating on the Polygon blockchain. Unlike Kalshi, Polymarket is globally accessible and allows pseudonymous trading via blockchain wallets. These structural differences make Polymarket useful for distinguishing patterns that appear exchange-specific from patterns that are also visible in a different institutional setting.

The earliest observed Kalshi trade timestamp in the local data is 1 July 2021, and the earliest observed Polymarket trade timestamp in the unified comparison data is 6 March 2023. The analysis applies a cutoff date of 31 December 2025. The Kalshi dataset comprises 64.7 million trades across 210,608 binary contracts, representing approximately 16.8 billion contracts traded. For each trade, the dataset includes the contract identifier, execution price in cents, number of contracts, the side taken by the trade initiator and the execution timestamp. For each market, the dataset includes the contract identifier, event category, resolution status, outcome and close time. Of the 210,608 markets, 98.6\% of those past their close date have resolved with a definitive yes/no outcome (Table~\ref{tab:kalshi_summary}). The Polymarket dataset comprises 288.7 million trades across 218,000 resolved contracts (79.8 billion contracts traded); the deposited unified snapshot is restricted to resolved contracts, since resolution outcomes are required to estimate calibration. Polymarket trade timestamps are derived from Polygon block numbers with approximately 3-hour noise, a limitation that affects the two shortest time bins (Section~\ref{sec:landscape}).

\subsection{Definitions}\label{sec:definitions}

\begin{table}[!t]
\caption{Terminology used throughout the analysis.\label{tab:definitions}}
\centering
\begin{tabular}{lp{0.66\columnwidth}}
\toprule
Term & Definition \\
\midrule
Contract & A binary yes/no claim paying \$1 if the stated event occurs and \$0 otherwise. \\
Market/event & The event-level grouping to which one or more contracts belong. Kalshi identifies events through ticker structure; Polymarket uses market titles and identifiers. \\
Trade & An executed transaction in a contract. The analysis uses the execution price, timestamp, number of contracts and, where available, the initiator side. \\
Execution price & The transaction price converted to the yes-claim price in cents. A price of 70 is commonly read as a 70\% probability before recalibration. \\
Trade size & The number of contracts in the executed transaction. Size bins are Single, Small, Medium and Large. \\
Domain & The event category used in the analysis: Sports, Politics, Crypto, Finance, Weather or Entertainment. \\
Resolution status & Whether the contract has a definitive yes/no outcome. Only definitively resolved contracts enter calibration estimates. \\
Time-to-resolution & The time between trade execution and contract close, discretized into the horizon bins in Section~\ref{sec:dimensions}. \\
Position limit & A platform rule limiting the exposure a participant may hold in a contract or event. Position limits are discussed as institutional context, not directly observed trader-level constraints. \\
\bottomrule
\end{tabular}
\end{table}
\subsection{Domain classification}\label{sec:domain}

Kalshi organizes contracts into events via a hierarchical ticker structure. Markets are classified into six knowledge domains (see supplementary material for full classification rules): \emph{Sports} (professional leagues including NFL, NBA, MLB and NHL), \emph{Politics} (elections, electoral college outcomes, government policy), \emph{Crypto} (cryptocurrency price contracts), \emph{Finance} (equity indices, interest rates, economic indicators), \emph{Weather} (temperature records, precipitation, natural events) and \emph{Entertainment} (awards, media, culture). The classification uses a deterministic mapping from event ticker prefixes, ensuring reproducibility. For Polymarket, which has no structured ticker namespace, markets are classified using compiled regular expression patterns applied to market titles. This yields three comparable domains (Sports, Crypto, Politics); Finance is excluded due to thin coverage (2,648 Polymarket markets vs 38,058 on Kalshi), and Weather and Entertainment have negligible Polymarket presence. The 42.5\% of Polymarket markets classified as `Other' reflects its long tail of bespoke markets (celebrity events, technology launches, meme markets). Tables~\ref{tab:kalshi_summary} and~\ref{tab:poly_summary} provide summary statistics.

\begin{table}[!t]
\caption{Summary statistics by domain: Kalshi (cutoff 31 December 2025).\label{tab:kalshi_summary}}
\centering
\resizebox{\columnwidth}{!}{
\begin{tabular}{lrrrrrr}
\toprule
Domain & Markets & Trades & Contracts & Resolved & Med.\ vol.\ & Base rate \\
\midrule
Sports        & 55,637   & 43.2M  & 12.7B  & 98.1\% & 76  & 41.3\% \\
Crypto        & 76,181   & 6.5M   & 742.9M & 99.1\% & 35  & 40.7\% \\
Politics      & 6,609    & 4.9M   & 2.2B   & 94.1\% & 127 & 40.2\% \\
Finance       & 38,058   & 4.3M   & 677.2M & 99.0\% & 38  & 37.7\% \\
Weather       & 26,911   & 4.4M   & 279.1M & 99.5\% & 74  & 24.0\% \\
Entertainment & 7,212    & 1.5M   & 174.5M & 96.7\% & 60  & 38.0\% \\
\midrule
\textbf{Total}& \textbf{210,608} & \textbf{64.7M} & \textbf{16.8B} & \textbf{98.6\%} & \textbf{47} & \textbf{38.1\%} \\
\bottomrule
\end{tabular}
}

\smallskip\noindent\emph{Note:} Resolved (\%) computed over markets past their close date. Base rate is the percentage of resolved markets where outcome = yes. Median volume is the median number of trades per market. Components may not sum to totals due to rounding.
\end{table}

Several features merit comment. Sports dominates Kalshi by volume (43.2 million trades, 66.7\% of the total) but Politics commands disproportionate contract value (2.2 billion contracts from only 6,609 markets, representing 13.3\% of total volume from 3.1\% of markets). Politics also has the highest median volume per market (127 trades), reflecting intense engagement with a smaller number of high-profile events. The distribution of trade sizes is heavily right-skewed across all domains: the median trade involves 40 contracts, but 0.15\% of trades (those exceeding 10,000 contracts) account for approximately 15\% of total contract volume. Polymarket (Table~\ref{tab:poly_summary}) shows a strikingly different composition: across the three comparable domains it has $3.4\times$ more trades than Kalshi from only $0.6\times$ as many markets, reflecting roughly $8\times$ higher median per-market volume. The most consequential difference is in Politics, where Polymarket has $9.3\times$ more trades (45.7M vs 4.9M) and $11\times$ more contracts (24.6B vs 2.2B), suggesting substantially deeper price discovery.

\begin{table}[!t]
\caption{Summary statistics by domain: Polymarket (three comparable domains; cutoff 31 December 2025).\label{tab:poly_summary}}
\centering
\resizebox{\columnwidth}{!}{
\begin{tabular}{lrrrrrr}
\toprule
Domain & Markets & Trades & Contracts & Resolved & Med.\ vol.\ & Base rate \\
\midrule
Sports   & 31,636  & 49.1M  & 21.1B  & 100\% & 64  & 35.3\% \\
Crypto   & 78,856  & 124.6M & 10.7B  & 100\% & 591 & 44.3\% \\
Politics & 14,389  & 45.7M  & 24.6B  & 100\% & 443 & 30.7\% \\
\midrule
\textbf{Total} & \textbf{124,881} & \textbf{219.4M} & \textbf{56.4B} & --- & \textbf{390} & \textbf{40.4\%} \\
\bottomrule
\end{tabular}
}

\smallskip\noindent\emph{Note:} Polymarket comparison is restricted to three domains with sufficient coverage; Finance is excluded due to thin coverage (2,648 markets). Weather and Entertainment have negligible Polymarket presence. The deposited unified Polymarket snapshot is restricted to resolved contracts (resolution outcomes are required to estimate calibration), so the resolved share is 100\% by construction. Median volume is the median number of trades per resolved market.
\end{table}

%% ============================================================
%% SECTION 3: MEASUREMENT
%% ============================================================
\section{Calibration measurement and empirical design}\label{sec:measuring}

\subsection{Logistic recalibration}\label{sec:logistic}

Calibration is measured using logistic recalibration \citep{platt1999probabilistic,niculescumizil2005predicting}. For a collection of $N$ trades indexed by $i$, each with market price $p_i \in (0,1)$ and binary outcome $y_i \in \{0,1\}$, the following logistic regression is fitted by maximum likelihood.
\begin{equation}\label{eq:logistic}
\mathrm{logit}\bigl(P(y_i = 1)\bigr) = a + b \cdot \mathrm{logit}(p_i),
\end{equation}
where $\mathrm{logit}(x) = \log\{x/(1-x)\}$. The log-likelihood is
\begin{equation}\label{eq:loglik}
\ell(a, b) = \sum_{i=1}^{N} \bigl[y_i \log \pi_i + (1 - y_i) \log(1 - \pi_i)\bigr],
\end{equation}
with $\pi_i = \sigma(a + b \cdot \mathrm{logit}(p_i))$ and $\sigma(x) = \{1 + \exp(-x)\}^{-1}$.
The intercept $a$ captures directional bias in yes/outcome frequencies, while the slope $b$ captures whether prices are too compressed or too extreme after accounting for that directional bias. When $a$ is close to zero, $b > 1$ corresponds to underconfidence: favorites are underpriced, longshots are overpriced and prices are compressed toward 50\%. Conversely, $b < 1$ corresponds to overconfidence, with prices that are too extreme. Because this interpretation is clearest when the intercept is not driving the result, the empirical tables retain both $a$ and $b$. The slope is used as the main decomposition statistic because it summarizes favorite--longshot compression, but the intercept estimates are reported in the supplementary calibration matrix and summarized in \ref{app:robustness}.

The analysis is restricted to trades with prices between 5 and 95 cents, excludes markets with fewer than 10 trades, requires at least 200 trades per analysis cell, and applies mild $L_2$ regularization ($C = 10$; \citealp{hosmer2013applied}). The price filter avoids near-certain contracts where tiny tick-size changes dominate the logit scale, while retaining the range in which probability interpretation is most common. The regularization stabilizes sparse cells and has negligible influence on the reported slopes: $C=1$, $C=10$ and $C=100$ give the same decomposition to three decimals (\ref{app:robustness}). All 216 Kalshi domain--horizon--size cells satisfy the sample-size requirement; domain-level minimum cell counts range from 472 trades in Weather to 22,518 trades in Sports.
\subsection{Analysis dimensions}\label{sec:dimensions}

For each trade, time-to-resolution is computed as $\tau = \text{close\_time} - \text{trade\_time}$ and discretized into nine bins, $[0, 1\text{h})$, $[1\text{h}, 3\text{h})$, $[3\text{h}, 6\text{h})$, $[6\text{h}, 12\text{h})$, $[12\text{h}, 24\text{h})$, $[24\text{h}, 48\text{h})$, $[2\text{d}, 1\text{w})$, $[1\text{w}, 1\text{mo})$ and $[1\text{mo}, \infty)$. Trade sizes are discretized into four bins, Single (1 contract), Small (2--10), Medium (11--100) and Large ($>$100). The full analysis grid comprises $6 \times 9 \times 4 = 216$ cells, all satisfying the minimum sample-size requirement. A total of 58.7 million Kalshi trades and 135.6 million Polymarket trades enter the calibration analysis after price filtering. Because Polymarket timestamps have approximately 3-hour noise (from block-number bucketing), the two shortest time bins (0--1h and 1--3h) are unreliable for cross-platform comparison; all Polymarket time-horizon results below use the seven reliable bins (3--6h through 1mo+).

As a functional-form check, the revision also computes binned reliability curves and isotonic calibration curves by domain from the Kalshi trade data. These nonparametric summaries are not used in the decomposition, but they provide a direct visual and tabular check on the logistic slope interpretation. For example, at a raw price of 0.75, the isotonic estimate is 0.886 in Politics and 0.691 in Weather, matching the slope-based conclusion that political prices are compressed while short-horizon weather prices tend to be too extreme. The generated reliability and isotonic tables are included in the replication output.
%% ============================================================
%% SECTION 3: THE CALIBRATION LANDSCAPE
%% ============================================================
\section{The calibration landscape}\label{sec:landscape}

\begin{figure}[!t]
\centering
\includegraphics[width=0.78\columnwidth]{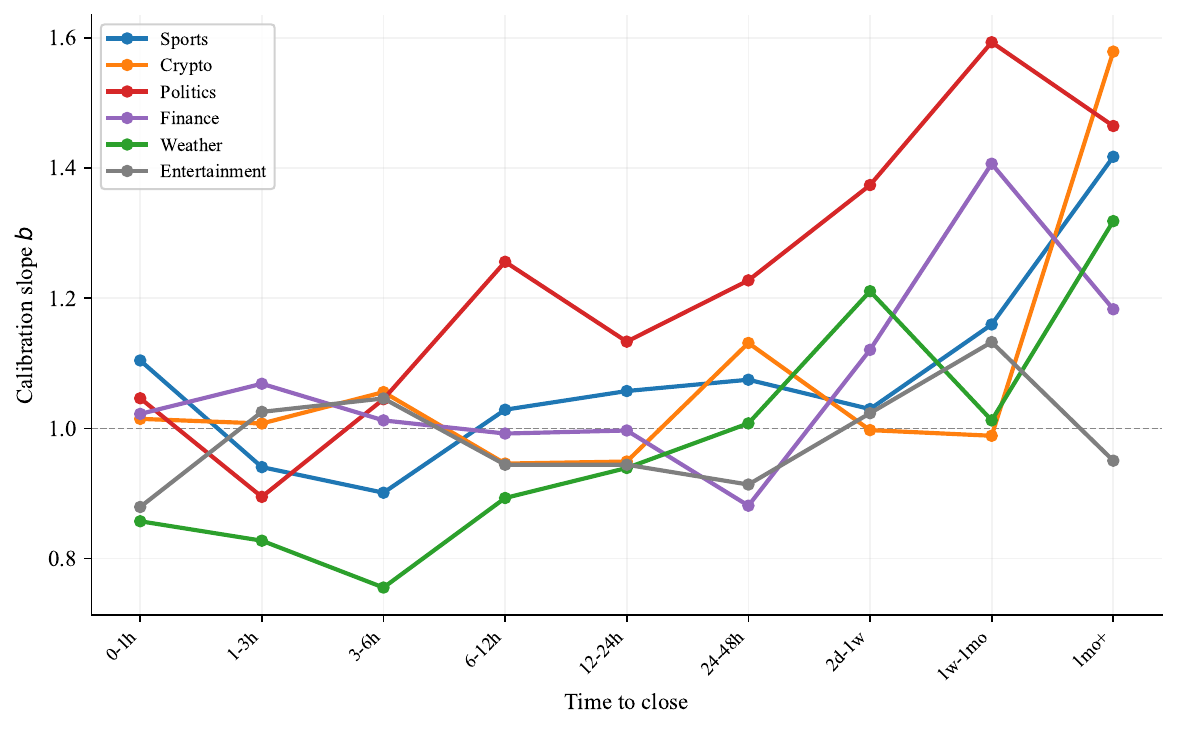}
\caption{Calibration slope $b$ versus time-to-resolution, one line per domain. These are pooled domain-by-time estimates matching Table~\ref{tab:slopes}; the decomposition in Section~\ref{sec:decomposition} uses the finer domain-by-time-by-size cell matrix. Slopes above 1 indicate underconfidence when intercepts are small relative to slope effects.}\label{fig:slope_trajectories}
\end{figure}

Figure~\ref{fig:slope_trajectories} plots calibration slopes against time-to-resolution for each domain. The detailed domain-by-time slopes are reported in Table~\ref{tab:slopes}. Both the figure and table use pooled domain-by-time estimates; the later decomposition uses the finer 216-cell domain-by-time-by-size matrix, so its horizon means are not simple averages of the six rows in Table~\ref{tab:slopes}. Slopes above 1 indicate underconfidence when intercepts are not driving the estimate (prices compressed toward 50\%); slopes below 1 indicate overconfidence (prices too extreme). Two features are immediately apparent: domains differ sharply, and these differences reflect different trajectory shapes, not mere level shifts.

\begin{figure}[!t]
\centering
\includegraphics[width=\columnwidth]{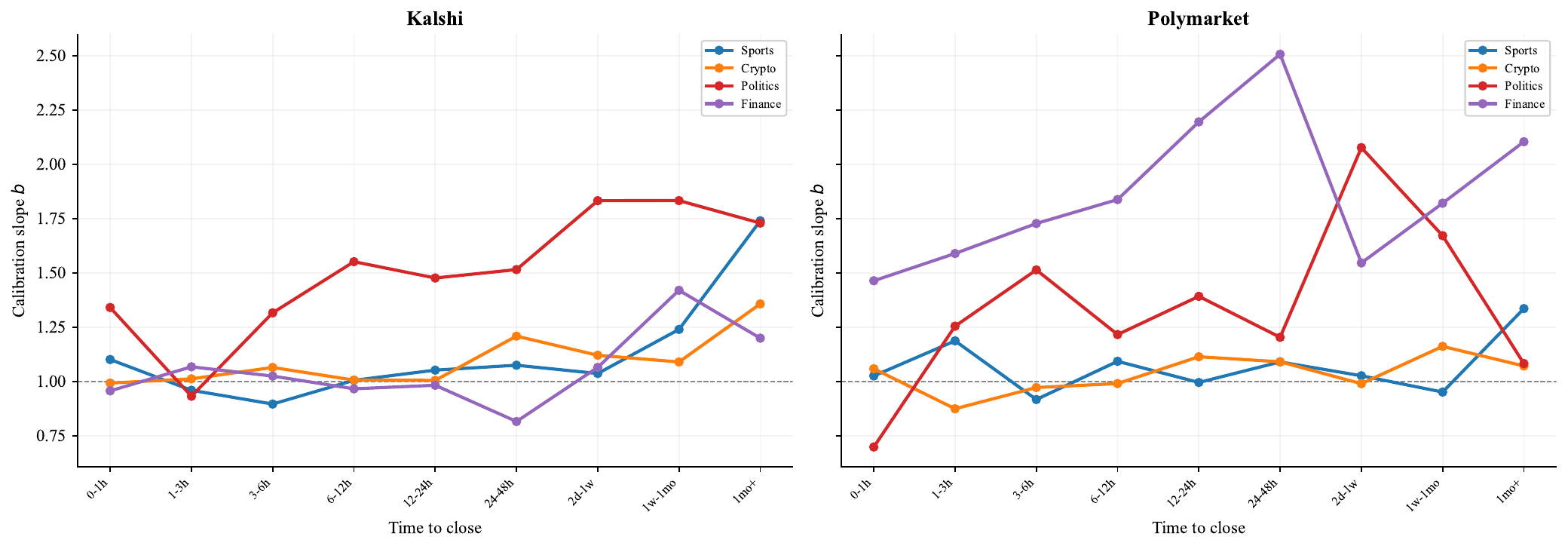}
\caption{Cross-platform calibration slope trajectories: (A) Kalshi and (B) Polymarket. The most stable cross-platform pattern is political underconfidence. Finance is shown on Polymarket for visual context but is excluded from formal cross-platform analysis due to thin coverage (2,648 markets).}\label{fig:cross_platform}
\end{figure}

\begin{table*}[!t]
\caption{Logistic recalibration slopes by domain and time-to-resolution. Values above 1.0 indicate underconfidence (prices compressed); below 1.0 indicate overconfidence (prices too extreme).\label{tab:slopes}}
\centering
\resizebox{\columnwidth}{!}{
\begin{tabular}{lccccccccc}
\toprule
Domain & 0--1h & 1--3h & 3--6h & 6--12h & 12--24h & 24--48h & 2d--1w & 1w--1mo & 1mo+ \\
\midrule
Politics      & 1.34 & 0.93 & 1.32 & 1.55 & 1.48 & 1.52 & 1.83 & 1.83 & 1.73 \\
Sports        & 1.10 & 0.96 & 0.90 & 1.01 & 1.05 & 1.08 & 1.04 & 1.24 & 1.74 \\
Crypto        & 0.99 & 1.01 & 1.07 & 1.01 & 1.01 & 1.21 & 1.12 & 1.09 & 1.36 \\
Finance       & 0.96 & 1.07 & 1.03 & 0.97 & 0.98 & 0.82 & 1.07 & 1.42 & 1.20 \\
Weather       & 0.69 & 0.84 & 0.73 & 0.87 & 0.91 & 0.97 & 1.20 & 1.20 & 1.37 \\
Entertainment & 0.81 & 1.02 & 1.00 & 0.92 & 0.89 & 0.84 & 1.07 & 1.11 & 0.96 \\
\bottomrule
\end{tabular}
}
\end{table*}
\subsection{Three stylised facts}\label{sec:facts}

\emph{Stylized Fact 1: Long-horizon prices are more compressed.} At long time horizons, prices in every domain move toward the favorite--longshot pattern: favorites are underpriced and longshots are overpriced. The universal horizon component $\mu(\tau)$, computed as the mean slope across all domain--size cells at each time bin (Section~\ref{sec:decomposition}), rises from 0.99 (0--1 hour) to 1.32 (beyond one month). This cell-level mean differs from the simple average of the six aggregate domain slopes in Table~\ref{tab:slopes}, which are contract-weighted pooled estimates. The pattern is consistent with prior evidence that favorite--longshot bias worsens with time to expiration \citep{page2013prediction} and is also visible on Polymarket across the three comparable domains.

\emph{Stylized Fact 2: Domains follow different calibration trajectories (the full $\beta$ matrix is in Table~\ref{tab:beta}).} Political markets exhibit persistent underconfidence at nearly all horizons (slopes 0.93--1.83), with prices compressed toward 50\%. As a slope-only illustration, a 70-cent political contract one week before resolution maps to approximately 83\% when applying the domain-horizon slope alone; Section~\ref{sec:practical} gives the intercept-aware recalibration formula. Sports markets are close to calibrated at short-to-medium horizons (slopes 0.90--1.10 from 0 to 48 hours) but become underconfident at long horizons, reaching 1.74 beyond one month. Weather markets exhibit the opposite pattern at short horizons (slopes 0.69--0.97 within 48 hours), where prices are too extreme. These domain-specific trajectories are also visible on Polymarket: Politics is underconfident (mean slope 1.45 across reliable bins), Sports is near-calibrated (1.06), and Crypto is mildly underconfident (1.06).

\emph{Stylized Fact 3: In political markets, large Kalshi trades are associated with greater price compression.} On Kalshi, large trades (over 100 contracts) in political markets produce calibration slopes of 1.74, compared to 1.19 for single-contract trades (Table~\ref{tab:scale}). The gap of 0.53 has a 95\% trade-level bootstrap confidence interval of $[0.29, 0.75]$ and remains positive under market-clustered ($[0.14, 1.32]$) and event-clustered ($[0.12, 1.80]$) resampling (Section~\ref{sec:significance}). In sports markets, the corresponding gap is 0.07 ($[-0.07, 0.26]$) and null under clustering. Because transaction prices are quantity-weighted in market data, large compressed trades have disproportionate influence on contract-weighted calibration estimates. The effect appears specific to Kalshi: on Polymarket the Politics gap is positive at the cell level ($\Delta = 0.28$, $[0.03, 0.54]$) but not robust once trades are clustered by market ($\Delta = 0.21$, $[-0.31, 1.12]$), and trade sizes are comparable across platforms (Politics median 43.5 vs 45 contracts), so the difference is not explained by larger Kalshi bets (Section~\ref{sec:institutions}).

\begin{table}[!t]
\caption{Calibration slopes by domain and trade size (Kalshi). $\Delta$(L$-$S) is the mean of time-bin-specific slope differences between the Large and Single bins (the bootstrap estimand). On Polymarket the Politics cell-level $\Delta$ is $+0.28$ $[0.03, 0.54]$ but is not robust to market clustering ($[-0.31, 1.12]$); Sports and Crypto are null (Table~\ref{tab:cross_bootstrap}).\label{tab:scale}}
\centering
\begin{tabular}{lccccc}
\toprule
Domain & Single & Small & Medium & Large & $\Delta$(L$-$S) \\
\midrule
Politics      & 1.19 & 1.22 & 1.37 & 1.74 & $+0.53$ \\
Sports        & 1.00 & 1.01 & 1.01 & 1.01 & $+0.07$ \\
Crypto        & 1.03 & 1.02 & 1.02 & 1.00 & $-0.02$ \\
Finance       & 1.10 & 1.08 & 1.04 & 1.05 & $-0.05$ \\
Weather       & 0.96 & 0.94 & 0.91 & 0.89 & $-0.07$ \\
Entertainment & 0.98 & 1.02 & 1.00 & 0.99 & $+0.01$ \\
\bottomrule
\end{tabular}
\end{table}
%% ============================================================
%% SECTION 4: DIAGNOSING POTENTIAL ARTEFACTS
%% ============================================================
\section{Diagnosing potential artifacts}\label{sec:artefacts}

\subsection{Is political underconfidence driven by a subset of markets?}\label{sec:subset}

Political markets encompass diverse subcategories (for example Electoral College, Governor, NYC Mayor and presidential-administration contracts). If underconfidence were confined to one subcategory, the conclusion would apply only to that subcategory rather than to political markets generally. Examining 10 political subcategories with sufficient data (see supplementary material), underconfidence appears in most of them rather than in one. Electoral College contracts show the strongest underconfidence (slopes 1.53--2.87 across time bins), while Trump Administration contracts span a wider range (0.54--1.64). Other Politics (1.42--2.38), Governor (1.19--4.02) and NYC Mayor (1.12--3.18) all exhibit persistent underconfidence.
\subsection{Composition effects in the 1--3 hour bin}\label{sec:composition}

Table~\ref{tab:slopes} shows that Politics achieves its lowest slope (0.93) at the 1--3 hour horizon, the only time bin where Politics appears slightly overconfident rather than underconfident. Disaggregation reveals a composition effect. Trump Administration markets comprise 63\% of trades in this bin and have a moderate slope of 1.08. Meanwhile, Electoral College contracts (7.4\% of trades, slope 1.81) and Other Politics (5.8\%, slope 2.17) remain strongly underconfident but are diluted by the Trump Administration majority. Leave-one-out analysis is consistent with this composition effect (a Simpson's-paradox pattern in which a heterogeneous mix of subcategories averages to a misleading aggregate): removing Trump Administration (the moderate majority) drops the aggregate from 0.93 to 0.88, revealing that the remaining subcategories are heterogeneous, with some strongly underconfident (Electoral College 1.81, Other Politics 2.17) and others overconfident (Biden Administration $-0.14$, a negative slope based on only 3.6\% of trades in this bin, likely reflecting thin, stale markets during the administration transition). Their opposing biases average to a misleadingly low aggregate. Removing Electoral College (the most underconfident minority) drops the aggregate to 0.69. The low aggregate slope reflects this subcategory composition, not a genuine regime shift toward overconfidence.
\subsection{Weighting sensitivity and the scale effect}\label{sec:weighting}

Prediction market prices implicitly weight traders by position size. To disentangle the market price (contract-weighted) from the crowd belief distribution (trade-weighted), the two weighting schemes are compared. In Politics, contract-weighted slopes exceed trade-weighted slopes by an average of 0.33 across all time bins, with the gap reaching 0.54 at the 2-day-to-1-week horizon. Contract-weighted prices are \emph{more} compressed than trade-weighted prices: larger positions are associated with prices further from truth. In Sports, the corresponding gap averages 0.06 and is concentrated at the longest horizon (0.35 at 1 month+, near zero elsewhere). The influence of large trades is amplified by the heavy right skew in trade sizes: the median trade involves 40 contracts, but just 0.15\% of trades (those exceeding 10,000 contracts) account for approximately 15\% of total contract volume, showing that a small number of large trades exert disproportionate influence on contract-weighted prices. On Polymarket, the Politics weighting gap is much smaller ($+0.08$) and oscillates in sign across time bins, consistent with the scale effect being largely specific to Kalshi.
\subsection{Nonparametric validation}\label{sec:nonparam}

\begin{figure}[!t]
\centering
\includegraphics[width=\columnwidth]{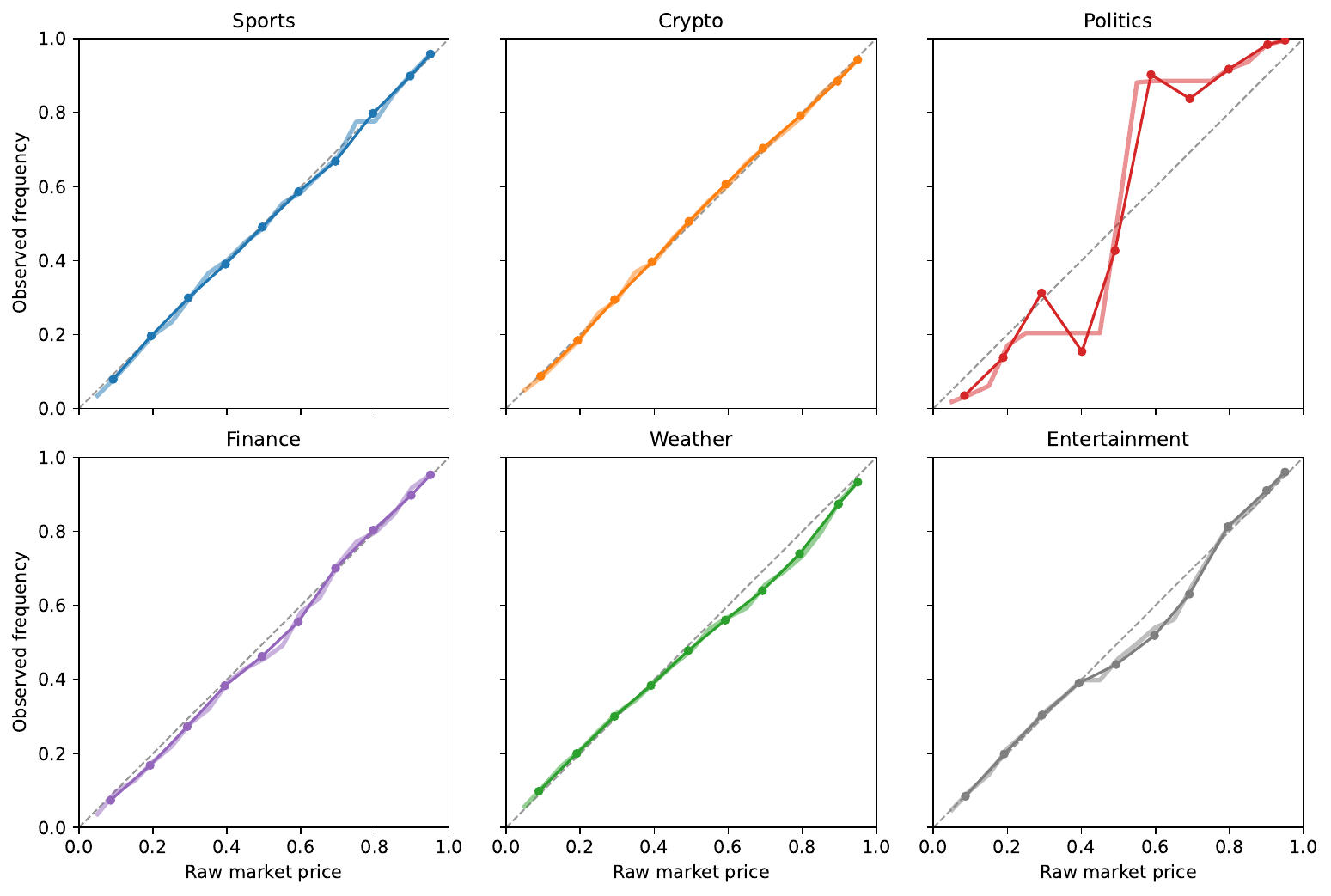}
\caption{Nonparametric reliability curves by domain: binned observed frequency against raw market price (markers) with the isotonic fit (shaded line) and the $45^\circ$ perfect-calibration line. Political prices are systematically compressed toward 50\%; short-horizon weather is mildly over-extreme; other domains track the diagonal.}\label{fig:reliability}
\end{figure}

Because the analysis summarizes calibration through a parametric (logistic) slope, this section verifies that the central findings do not depend on that functional form. For each domain, fully nonparametric calibration estimators are computed on the same trade data: binned reliability curves, an isotonic fit, and the expected and maximum calibration error (ECE, MCE) together with the Murphy decomposition of the Brier score into reliability, resolution and uncertainty (Table~\ref{tab:nonparam}; Figure~\ref{fig:reliability}). The model-free picture matches the slope-based one and is, if anything, starker: Politics has an ECE of 0.117 and a reliability (miscalibration) component of 0.024---five to fifteen times any other domain, all of which sit near zero---and ranking domains by ECE places Politics far worst---matching the slope-based conclusion---while the remaining, well-calibrated domains cluster near zero (Sports and Crypto lowest). The isotonic curves show the same shapes the slopes imply: political prices compressed toward 50\%, short-horizon weather mildly over-extreme, and the remaining domains close to the diagonal. The slope is therefore a faithful low-dimensional summary, chosen for parsimony and because it supports the additive decomposition, not because it manufactures the result.

\begin{table}[!t]
\caption{Nonparametric calibration metrics by domain (contract-weighted). ECE and MCE are the expected and maximum calibration error over price deciles; the Brier score decomposes as reliability $-$ resolution $+$ uncertainty (Murphy). Lower ECE/MCE/reliability indicate better calibration.\label{tab:nonparam}}
\centering
\begin{tabular}{lccccc}
\toprule
Domain & ECE & MCE & Brier & Reliability & Resolution \\
\midrule
Politics      & 0.117 & 0.315 & 0.119 & 0.024 & 0.147 \\
Entertainment & 0.022 & 0.079 & 0.160 & 0.001 & 0.087 \\
Finance       & 0.016 & 0.037 & 0.156 & 0.000 & 0.083 \\
Weather       & 0.016 & 0.053 & 0.172 & 0.000 & 0.057 \\
Sports        & 0.008 & 0.025 & 0.185 & 0.000 & 0.063 \\
Crypto        & 0.007 & 0.013 & 0.174 & 0.000 & 0.073 \\
\bottomrule
\end{tabular}
\end{table}

%% ============================================================
%% SECTION 5: THE DECOMPOSITION MODEL
%% ============================================================
\section{The decomposition model}\label{sec:decomposition}

\subsection{Framework}\label{sec:framework}

The patterns in Section~\ref{sec:landscape} suggest an additive descriptive structure. The calibration slope is modeled as $\theta$ observed in cell $(d, \tau, s)$, where $d$ indexes domain, $\tau$ indexes time-to-resolution bin and $s$ indexes trade-size bin:
\begin{equation}\label{eq:decomp}
\theta(d, \tau, s) = \mu(\tau) + \alpha_d + \kappa(s) + \beta_d(\tau) + \gamma_d(s) + \varepsilon,
\end{equation}
where $\mu(\tau)$ is a \emph{universal horizon function} capturing the shared tendency toward underconfidence at long horizons; $\alpha_d$ is a \emph{domain intercept} capturing persistent bias; $\kappa(s)$ is a common trade-size main effect; $\beta_d(\tau)$ is a \emph{domain-by-horizon interaction}; and $\gamma_d(s)$ is a \emph{domain-by-size interaction} capturing how trade size modulates calibration differently across domains after the common size effect is removed. The model is identified by the following constraints.
\begin{align}
\sum_{d=1}^{D} \alpha_d &= 0, \label{eq:id1}\\
\sum_{d=1}^{D} \beta_d(\tau) = 0 \;\; \forall\, \tau, \quad &\sum_{\tau} \beta_d(\tau) = 0 \;\; \forall\, d, \label{eq:id2}\\
\sum_{s=1}^{S} \kappa(s) &= 0, \label{eq:id3}\\
\sum_{d=1}^{D} \gamma_d(s) = 0 \;\; \forall\, s, \quad &\sum_{s=1}^{S} \gamma_d(s) = 0 \;\; \forall\, d. \label{eq:id4}
\end{align}
These ensure that the universal horizon function $\mu(\tau)$ absorbs the cross-domain mean at each horizon, the domain intercepts $\alpha_d$ absorb the cross-horizon mean for each domain, and the common size effect is centered. The interaction terms $\beta_d(\tau)$ and $\gamma_d(s)$ are \emph{doubly centered} (over domains and over the horizon or size index, respectively), which is why they carry $(D-1)(T-1)=40$ and $(D-1)(S-1)=15$ degrees of freedom in the $72$-parameter count.

The additive specification is chosen for interpretability and parsimony. In logit space, calibration slopes are unbounded and approximately continuous, making additive decomposition a useful first-order summary. A multiplicative specification (modeling log slopes) was also considered; the two yield nearly identical fits because most slopes are concentrated in the range $[0.7, 1.9]$. A non-domain-specific size--horizon interaction term was tested (Section~\ref{sec:confound}) and found to explain only 2.6\% additional variance, insufficient to justify the 54 additional parameters it requires. The Bayesian hierarchical model in Section~\ref{sec:bayesian} provides a complementary uncertainty-aware check, but the decomposition itself should be read as descriptive rather than causal or structural.
\subsection{Variance decomposition}\label{sec:variance}

The decomposition is estimated using sequential projection, analogous to Type~I sums of squares. Let $\bar{\theta}$ denote the grand mean of all 216 observed slopes. The total sum of squares is
\begin{equation}\label{eq:sstot}
\mathrm{SS}_{\mathrm{tot}} = \sum_{d,\tau,s} \bigl[\theta(d,\tau,s) - \bar{\theta}\bigr]^2,
\end{equation}
and the residual sum of squares after fitting~\eqref{eq:decomp} is
\begin{equation}\label{eq:ssres}
\mathrm{SS}_{\mathrm{res}} = \sum_{d,\tau,s} \bigl[\theta(d,\tau,s) - \hat{\theta}(d,\tau,s)\bigr]^2,
\end{equation}
where $\hat{\theta}(d,\tau,s) = \hat{\mu}(\tau) + \hat{\alpha}_d + \hat{\kappa}(s) + \hat{\beta}_d(\tau) + \hat{\gamma}_d(s)$. Two quantities are reported per component. The \emph{marginal} $R^2$ is the increment in explained variance when that component enters the sequential fit ($R^2_\mu = 1 - \mathrm{SS}_{\mathrm{res}}^{(\mu)} / \mathrm{SS}_{\mathrm{tot}}$, then $R^2_\alpha = R^2_{\mu,\alpha} - R^2_\mu$, and so on); the \emph{cumulative} $R^2$ is the running total. Because this is a Type~I (sequential) decomposition, the marginal shares depend on the entry order, fixed here as horizon, domain, size, domain~$\times$~horizon, domain~$\times$~size; the order dependence is examined with Type~II and Type~III sums of squares below. Table~\ref{tab:variance} reports the variance explained. The model accounts for 87.3\% of calibration variance across 216 cells, with 72 fitted degrees of freedom (9 for $\mu$, 5 for $\alpha$, 3 for $\kappa$, 40 for $\beta$ and 15 for $\gamma$), yielding an adjusted $R^2$ of 0.810.

Two checks address the concern that this in-sample $R^2$ might reflect the flexibility of a 72-parameter model rather than genuine structure. First, a permutation null in which the slope vector is randomly reassigned across cells and the full model refitted gives a mean null $R^2$ of only 0.329 (95th percentile 0.406, maximum 0.531 over 5{,}000 permutations); the observed 0.873 lies far outside this distribution ($p < 0.001$). A model of this size therefore explains roughly a third of the variance by chance alone, and the observed fit is well above that floor. Second, out-of-sample predictive accuracy is substantial: leave-one-cell-out $R^2 = 0.715$ and 10-fold cross-validated $R^2 = 0.704$. More demanding leave-one-domain or leave-one-size-bin checks perform poorly, so the decomposition summarizes the observed grid rather than extrapolating to unobserved domains or size regimes. Section~\ref{sec:bayesian} complements these with an uncertainty-aware estimate that propagates first-stage error.

\begin{table}[!t]
\caption{Variance decomposition of calibration slopes (216 cells, Type~I order: horizon, domain, size, domain $\times$ horizon, domain $\times$ size).\label{tab:variance}}
\centering
\resizebox{\columnwidth}{!}{
\begin{tabular}{llcc}
\toprule
Component & Interpretation & Marginal $R^2$ & Cumulative $R^2$ \\
\midrule
$\mu(\tau)$       & Universal horizon effect       & 0.302 & 0.302 \\
$\alpha_d$        & Domain intercept               & 0.146 & 0.448 \\
$\kappa(s)$       & Common size effect             & 0.032 & 0.481 \\
$\beta_d(\tau)$   & Domain $\times$ horizon interaction & 0.260 & 0.741 \\
$\gamma_d(s)$     & Domain $\times$ size interaction & 0.133 & 0.873 \\
$\varepsilon$     & Residual                       & 0.127 & --- \\
\bottomrule
\end{tabular}
}
\end{table}

The domain-by-horizon interaction $\beta$ is the single largest domain-specific component (26.0\%). Knowing \emph{which} domain a market belongs to and \emph{when} the price is observed is more informative about calibration quality than either piece of information alone. The common size effect is modest (3.2\%), but the remaining domain-by-size term is larger (13.3\%), which is why the size result is interpreted as domain-specific rather than a universal feature of large trades.

Because Type~I decomposition is order-dependent, verification is conducted with Type~II and Type~III sums of squares. The interaction terms are stable across decomposition orders, while the main effects shift as expected when time, domain and size share variance. Additionally, weighting cells by inverse estimation variance yields the weighted decomposition
\begin{equation}\label{eq:wls}
\hat{\boldsymbol{\phi}}_{\mathrm{WLS}} = \arg\min_{\boldsymbol{\phi}} \sum_{d,\tau,s} w_{d\tau s} \bigl[\theta(d,\tau,s) - \hat{\theta}(d,\tau,s;\, \boldsymbol{\phi})\bigr]^2, \quad w_{d\tau s} = 1/\mathrm{SE}^2_{d\tau s},
\end{equation}
where $\boldsymbol{\phi} = (\mu, \alpha, \kappa, \beta, \gamma)$ collects all model parameters. This yields a total weighted $R^2$ of 0.995, with the horizon effect $\mu$ dominating (0.74 vs 0.30 unweighted). This is an inverse-variance weighted \emph{in-sample} fit, not an out-of-sample measure; its high value indicates that the 12.7\% unweighted residual is concentrated in noisier, lower-volume cells rather than in precisely estimated ones. The out-of-sample picture is given by the leave-one-cell-out and $k$-fold checks above and by the measurement-error model of Section~\ref{sec:bayesian}. The domain intercepts are also compared on Polymarket: Politics remains the clear positive outlier (mean slope 1.45 at reliable horizons), while Sports (1.06) and Crypto (1.06) cluster near perfect calibration (Appendix~C reports full cross-platform comparison tables).

\begin{figure}[!t]
\centering
\includegraphics[width=\columnwidth]{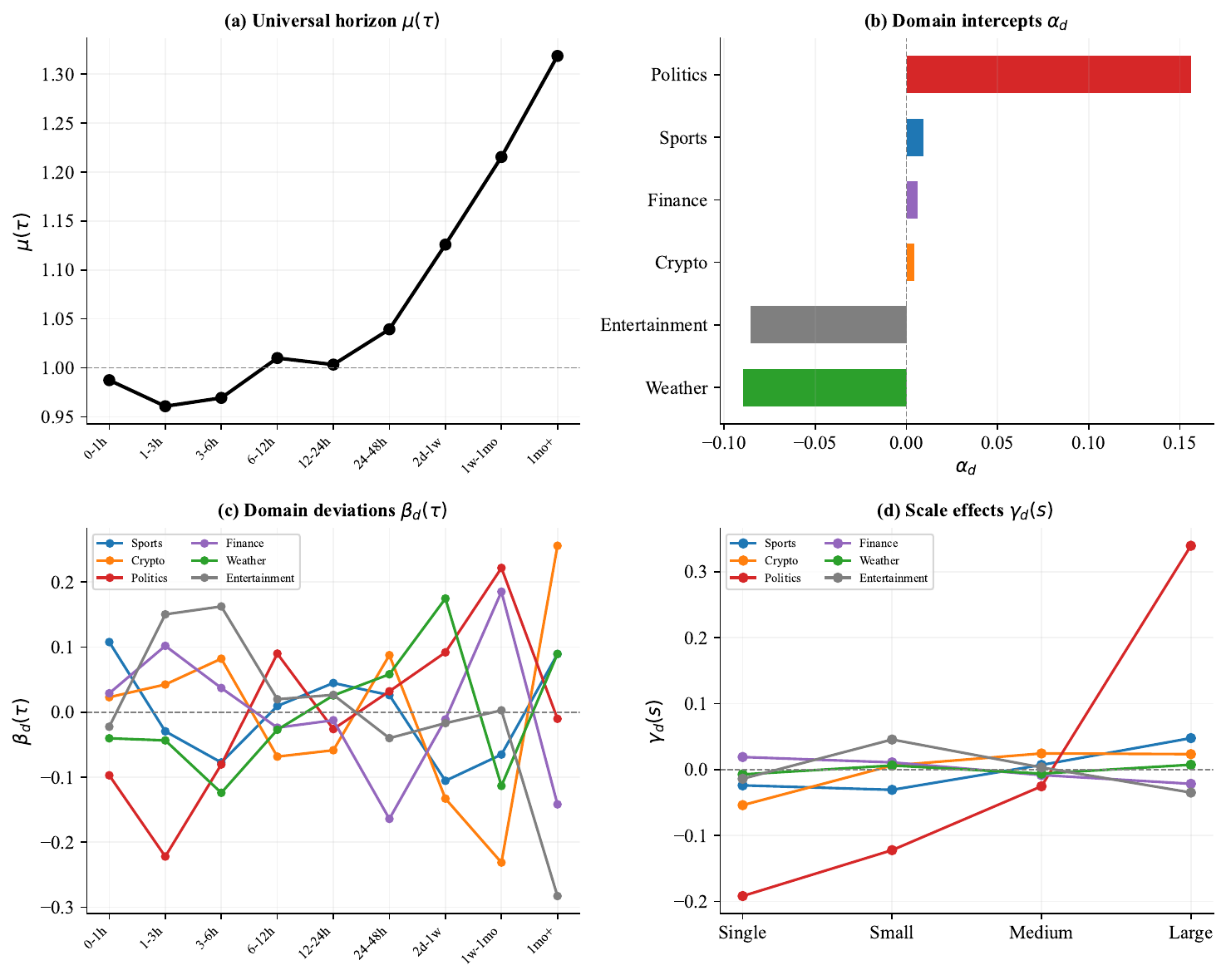}
\caption{Decomposition of calibration slopes. (A) Universal horizon effect $\mu(\tau)$. (B) Domain intercepts $\alpha_d$. (C) Domain-by-horizon interactions $\beta_d(\tau)$. (D) Size-related effects by domain; Table~\ref{tab:variance} separates the common size main effect from the domain-by-size interaction.}\label{fig:hero}
\end{figure}

\begin{figure}[!t]
\centering
\includegraphics[width=0.58\columnwidth]{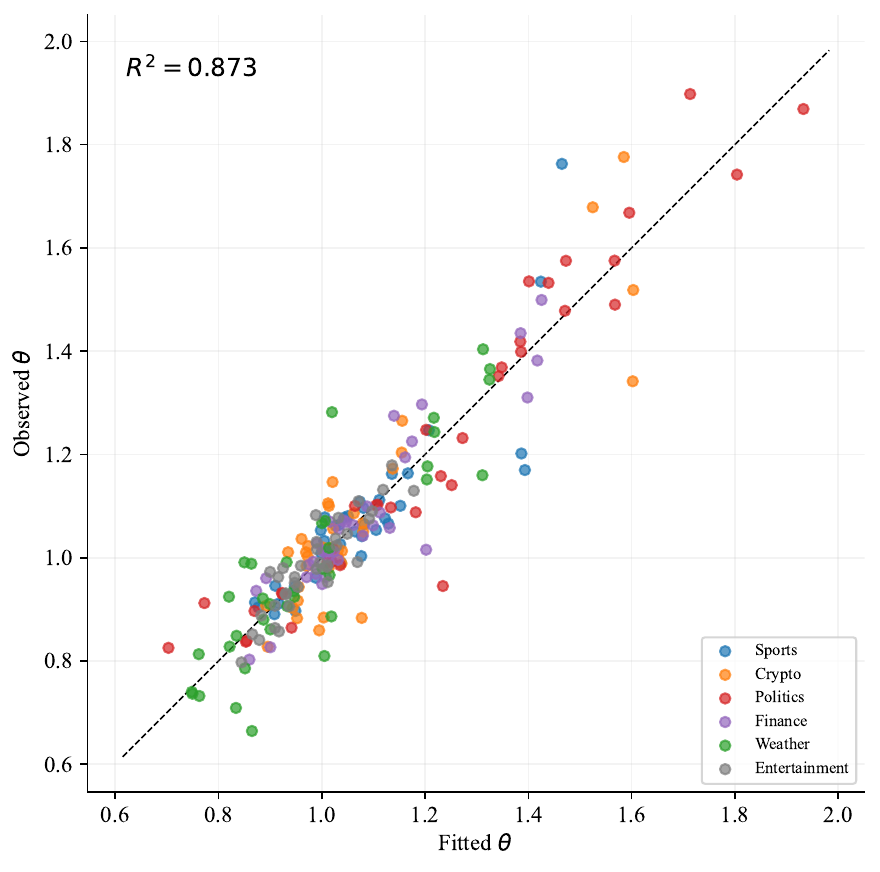}
\caption{Observed versus fitted calibration slopes ($R^2 = 0.873$). Each point is one of 216 analysis cells.}\label{fig:obs_vs_fit}
\end{figure}
\subsection{Statistical evidence}\label{sec:significance}

The main and interaction terms are statistically distinguishable from residual variation in the 216-cell matrix (Table~\ref{tab:ftest}). Domain intercepts differ ($F(5, 144) = 33.16$, $p < 10^{-16}$). The common size main effect is smaller but nonzero ($F(3, 144) = 12.24$, $p < 10^{-6}$). Domain-specific trajectories remain important ($F(40, 144) = 7.40$, $p < 10^{-16}$), and the residual domain-by-size interaction is also distinguishable from zero ($F(15, 144) = 10.05$, $p < 10^{-15}$).

The Politics scale effect (the slope difference of 0.53 between large and single-contract trades) is formally defined as
\begin{equation}\label{eq:delta}
\Delta_d = \frac{1}{T} \sum_{\tau=1}^{T} \bigl[\theta(d, \tau, s_{\text{L}}) - \theta(d, \tau, s_{\text{S}})\bigr],
\end{equation}
where $s_{\text{L}}$ and $s_{\text{S}}$ index the Large and Single trade-size bins respectively, and $T = 9$ is the number of time bins. This within-horizon averaging ensures that the estimand is not confounded by compositional differences across horizons. Because trades are not independent---many occur within the same contract, and the yes and no sides of an event are mechanically linked---the trade-level interval is not relied upon alone. Whole clusters are resampled at two levels: by market (contract) and by event (\texttt{event\_ticker}, which groups the related contracts of one event). For Politics the effect remains positive and excludes zero under both: market-clustered $[0.14, 1.32]$ and event-clustered $[0.12, 1.80]$ (the event-clustered interval is widest, with only 850 clusters). The trade-level interval is $[0.29, 0.75]$. The Sports scale effect remains null under all three (trade-level $[-0.07, 0.26]$; market-clustered $[-0.03, 0.05]$; event-clustered $[-0.03, 0.06]$). The same event-clustered standard errors feed the measurement-error model of Section~\ref{sec:bayesian}; they are on the order of 50 times the naive Fisher errors that assume independent trades, which is why the descriptive precision is interpreted cautiously.
\subsection{Is the scale effect confounded with time horizon?}\label{sec:confound}

If large trades concentrate at long horizons, the scale effect could be an artifact of horizon confounding. Three checks make this less likely. First, in Politics, large trades have \emph{shorter} median horizons (213 hours) than single-contract trades (862 hours); confounding would therefore bias \emph{against} finding a scale effect, not in favor of it. Second, adding a non-domain-specific size--horizon interaction to the model explains only 2.6\% additional variance, while the domain-specific size interaction remains 13.3\% after the common size main effect is separated. Third, within Politics, the scale effect is positive in all nine time bins, ranging from $+0.12$ (1--3 hours) to $+0.81$ (2 days--1 week). The scale effect is therefore unlikely to be only a horizon artifact (Tables~\ref{tab:confound} and~\ref{tab:ftest}).

\begin{figure}[!t]
\centering
\includegraphics[width=0.78\columnwidth]{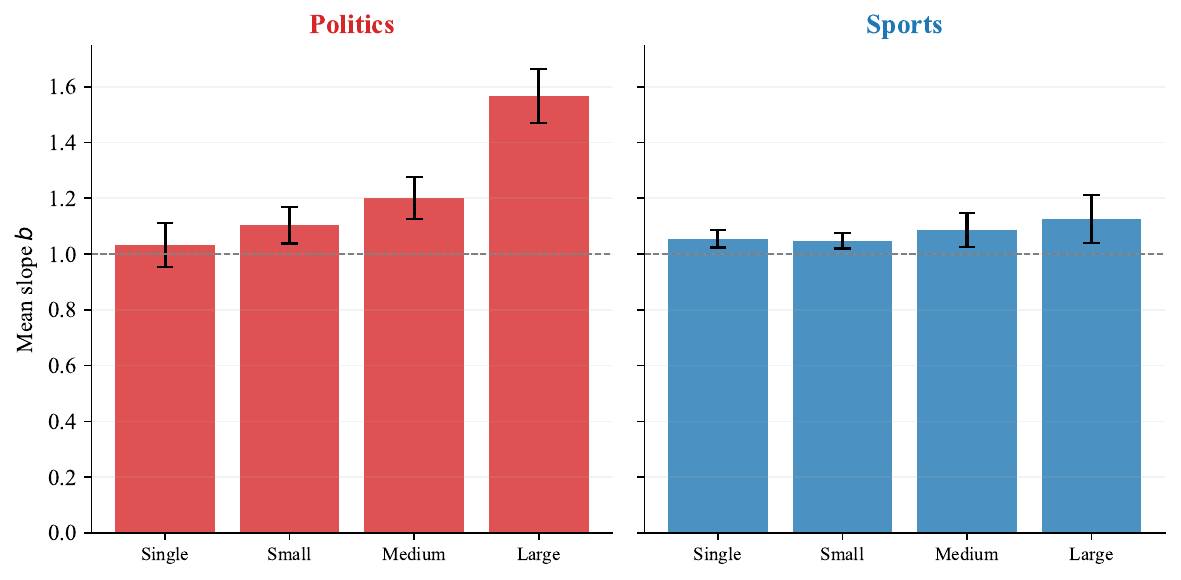}
\caption{Calibration slopes by trade size: (A) Politics versus (B) Sports. In Politics, larger trades are associated with more compressed prices (higher slopes). In Sports, no such gradient exists.}\label{fig:whale}
\end{figure}

\begin{figure}[!t]
\centering
\includegraphics[width=0.78\columnwidth]{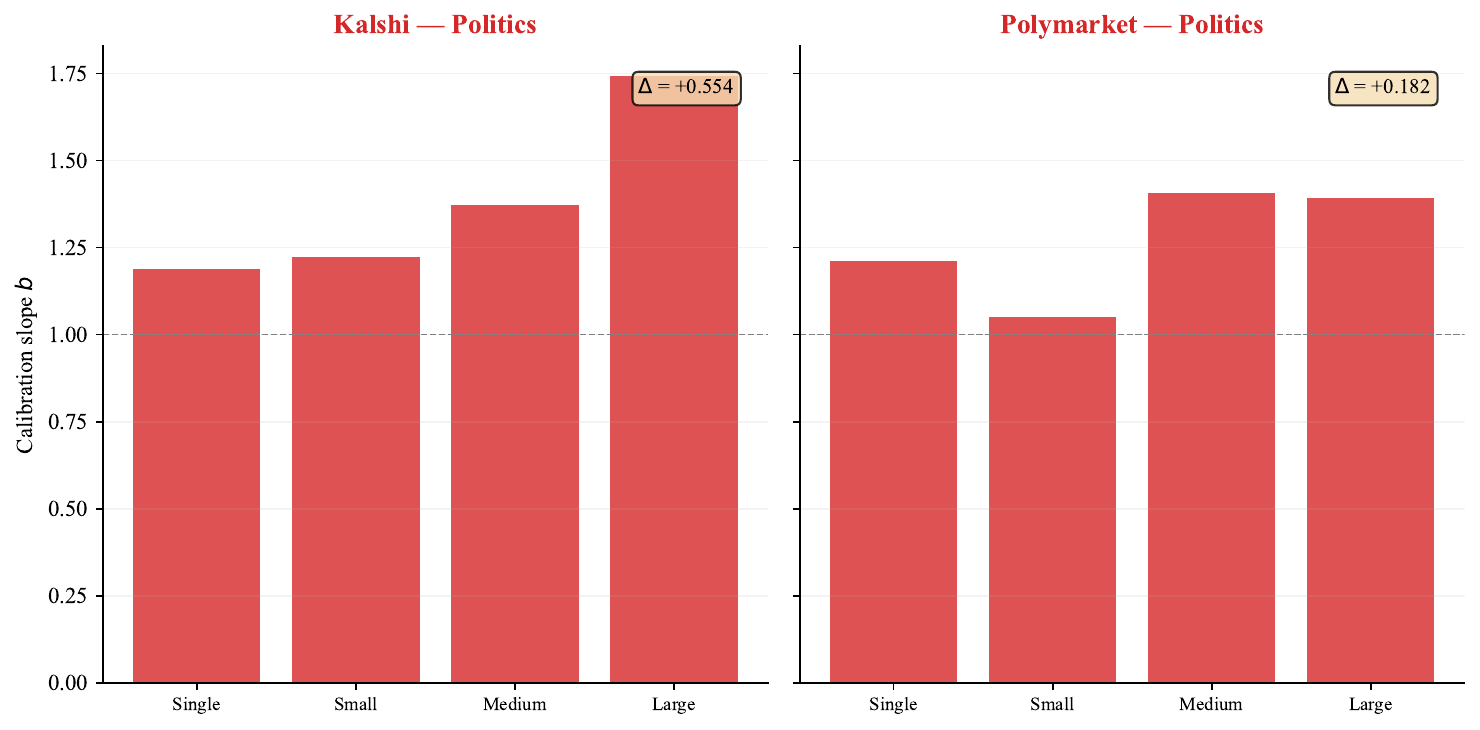}
\caption{Cross-platform scale effect in Politics. (A) Kalshi shows a monotonic increase from Single (1.19) to Large (1.74), $\Delta = +0.53$ $[0.29, 0.75]$, robust to clustering. (B) Polymarket shows a non-monotonic pattern that is positive at the cell level ($\Delta = +0.28$ $[0.03, 0.54]$) but not robust to market clustering. Bar labels show aggregate slope differences (Table~\ref{tab:cross_scale}); bootstrap CIs use the within-horizon estimand of Equation~\eqref{eq:delta}.}\label{fig:cross_scale}
\end{figure}
%% ============================================================
%% SECTION 6: BAYESIAN HIERARCHICAL MODEL
%% ============================================================
\section{Bayesian hierarchical model}\label{sec:bayesian}

The decomposition in Section~\ref{sec:decomposition} is descriptive: it treats each estimated slope as if it were observed without error. As one reviewer noted, this second stage models quantities that are themselves estimated, so estimation uncertainty should be propagated into the decomposition. This section develops a hierarchical \emph{measurement-error} model that carries each cell's first-stage standard error directly into the likelihood, providing uncertainty-aware inference and partial pooling.
\subsection{Model specification}\label{sec:bayesian_spec}

Let $\theta_{\mathrm{obs}}(d, \tau, s)$ denote the first-stage slope estimate in cell $(d, \tau, s)$ and $\mathrm{se}_{d\tau s}$ its event-clustered standard error (Section~\ref{sec:significance}; \ref{app:robustness}). Marginalising the latent true slope, the observation level is
\begin{equation}\label{eq:bayesian}
\theta_{\mathrm{obs}}(d, \tau, s) \sim \mathcal{N}\!\Bigl(\mu(\tau) + \alpha_d + \beta_d(\tau) + \delta_d \cdot \tilde{s},\; \sigma^2 + \mathrm{se}_{d\tau s}^2\Bigr),
\end{equation}
where $\tilde{s} = \log s - \overline{\log s}$ is the centered log trade size for bin $s$. The key feature is the variance term: $\mathrm{se}_{d\tau s}^2$ is the known first-stage (event-clustered) estimation variance, and $\sigma^2$ is \emph{residual structural} heterogeneity beyond measurement noise. Noisily estimated cells are therefore down-weighted automatically, and the decomposition no longer treats the slopes as exact. At the prior level, each component receives a hierarchical structure.
\begin{align}
\mu(\tau) &\sim \mathcal{N}(1.0,\; 0.5^2), \quad \tau = 1, \ldots, 9, \label{eq:prior_mu}\\
\alpha_d &\sim \mathcal{N}(0,\; \sigma_\alpha^2), \quad \sum_{d=1}^{6}\alpha_d = 0, \label{eq:prior_alpha}\\
\beta_d(\tau) &\sim \mathcal{N}(0,\; \sigma_\beta^2), \quad \sum_{d}\beta_d(\tau) = 0 \;\forall\, \tau, \label{eq:prior_beta}\\
\delta_d &\sim \mathcal{N}(0,\; \sigma_\delta^2). \label{eq:prior_delta}
\end{align}
The prior for $\mu(\tau)$ is centered at perfect calibration ($b = 1$). At the hyperprior level,
\begin{equation}\label{eq:hyperprior}
\sigma_\alpha,\; \sigma_\beta,\; \sigma_\delta,\; \sigma \;\overset{\mathrm{ind.}}{\sim}\; \mathrm{Half\text{-}Cauchy}(0,\, 1).
\end{equation}
The non-centered parameterization $\alpha_d = \sigma_\alpha \cdot \alpha_d^{\mathrm{raw}}$, with $\alpha_d^{\mathrm{raw}} \sim \mathcal{N}(0,1)$, is used for all hierarchical effects; $\beta_d(\tau)$ is doubly centered (over domains and over horizons). The scale effect uses the continuous specification $\delta_d \tilde{s}$ rather than the categorical $\gamma_d(s)$, so close agreement across the two parameterizations is itself a robustness check.
\subsection{Results}\label{sec:bayesian_results}

Posterior inference is conducted via Hamiltonian Monte Carlo \citep{neal2011mcmc,betancourt2017conceptual} using NumPyro \citep{phan2019composable} (4 chains, 4,000 iterations with 3,000 warmup, target acceptance 0.99). The maximum split-$\hat{R}$ \citep{vehtari2021rank,brooks1998general} is 1.000, the minimum bulk effective sample size is 3,794, and no divergent transitions occurred.

Table~\ref{tab:bayesian_intercepts} reports domain intercepts. The qualitative ranking agrees with the descriptive decomposition: Politics is the clear positive outlier (posterior mean $+0.107$, 95\% credible interval $[0.062, 0.152]$, entirely above zero, indicating persistent underconfidence), while Weather ($-0.072$, $[-0.111, -0.034]$) and Entertainment ($-0.062$, $[-0.096, -0.029]$) are below zero. Sports, Crypto and Finance are indistinguishable from zero. Because the model now propagates the (large) event-clustered first-stage uncertainty, the posterior means are pulled toward the common mean more than in the descriptive estimates---the Politics intercept shrinks from $+0.156$ to $+0.107$---which is the expected and intended effect of uncertainty-aware partial pooling rather than a contradiction.

\begin{table}[!t]
\caption{Bayesian posterior summaries for domain intercepts $\alpha_d$, from the measurement-error model with event-clustered first-stage standard errors.\label{tab:bayesian_intercepts}}
\centering
\begin{tabular}{lcccc}
\toprule
Domain & Post.\ mean & SD & 95\% CrI & Freq.\ \\
\midrule
Politics      & $+0.107$ & 0.023 & $[+0.062, +0.152]$ & $+0.156$ \\
Sports        & $+0.013$ & 0.016 & $[-0.019, +0.043]$ & $+0.009$ \\
Crypto        & $-0.004$ & 0.017 & $[-0.038, +0.030]$ & $+0.004$ \\
Finance       & $+0.019$ & 0.013 & $[-0.008, +0.045]$ & $+0.006$ \\
Weather       & $-0.072$ & 0.019 & $[-0.111, -0.034]$ & $-0.090$ \\
Entertainment & $-0.062$ & 0.017 & $[-0.096, -0.029]$ & $-0.086$ \\
\bottomrule
\end{tabular}

\smallskip\noindent\emph{Note:} CrI denotes a 95\% credible interval. The Freq.\ column reports the corresponding descriptive (frequentist) estimate from Table~\ref{tab:variance}. Differences reflect shrinkage induced by propagating first-stage uncertainty.
\end{table}

The scale sensitivity $\delta_d$ is positive and credibly different from zero only in Politics ($\delta = +0.058$, 95\% credible interval $[0.031, 0.085]$): larger trades are associated with more compressed prices in political markets and in no other domain, consistent with the frequentist scale effect. The residual structural standard deviation is small ($\sigma = 0.006$, 95\% credible interval $[0.000, 0.017]$), meaning that once the four structural components and measurement error are accounted for, little further cell-to-cell heterogeneity is detected.

This last point reframes the explanatory-power question raised in review. Treating the slopes as exact, the descriptive decomposition accounts for 87.3\% of cell-level slope variance (Section~\ref{sec:variance}). Once first-stage uncertainty is propagated, however, the event-clustered measurement variance accounts for roughly half of the raw cell-to-cell dispersion in slopes, and the four structural components explain $45.6\%$ of the total observed variance (95\% credible interval $[31\%, 62\%]$; Table~\ref{tab:variance_components}). The two figures are complementary rather than contradictory: 87.3\% is the share of \emph{observed} slope variation captured by the structure, while $\approx\!46\%$ is the share that is \emph{structural signal} rather than estimation noise. Because $\sigma \approx 0$, little residual heterogeneity is \emph{detected under this measurement-error specification}---which is a statement about this model, not a claim that no structure remains; what shrinks relative to the descriptive fit is the apparent precision, not the substantive pattern. One limitation should be stated plainly: the model propagates each cell's marginal first-stage variance but not the cross-cell covariance induced by events that contribute trades to several cells, so the reported credible intervals may still be mildly optimistic. The event-clustered bootstrap of the scale effect (Section~\ref{sec:significance}), which resamples whole events through the pipeline, is not subject to this limitation and supports the same conclusion.

\begin{table}[!t]
\caption{Posterior share of observed slope variance attributable to each structural component (measurement-error model). Shares are mutually orthogonal on the balanced grid; the remainder is first-stage estimation noise.\label{tab:variance_components}}
\centering
\begin{tabular}{lcc}
\toprule
Component & Posterior mean share & 95\% credible interval \\
\midrule
$\mu(\tau)$ horizon              & 0.193 & $[0.111, 0.294]$ \\
$\alpha_d$ domain                & 0.084 & $[0.039, 0.141]$ \\
$\beta_d(\tau)$ domain $\times$ horizon & 0.111 & $[0.058, 0.179]$ \\
$\delta_d\tilde{s}$ scale        & 0.068 & $[0.022, 0.129]$ \\
\midrule
Total structural                 & 0.456 & $[0.315, 0.620]$ \\
\bottomrule
\end{tabular}
\end{table}
\subsection{Posterior predictive check}\label{sec:ppc}

\begin{figure}[!t]
\centering
\includegraphics[width=0.68\columnwidth]{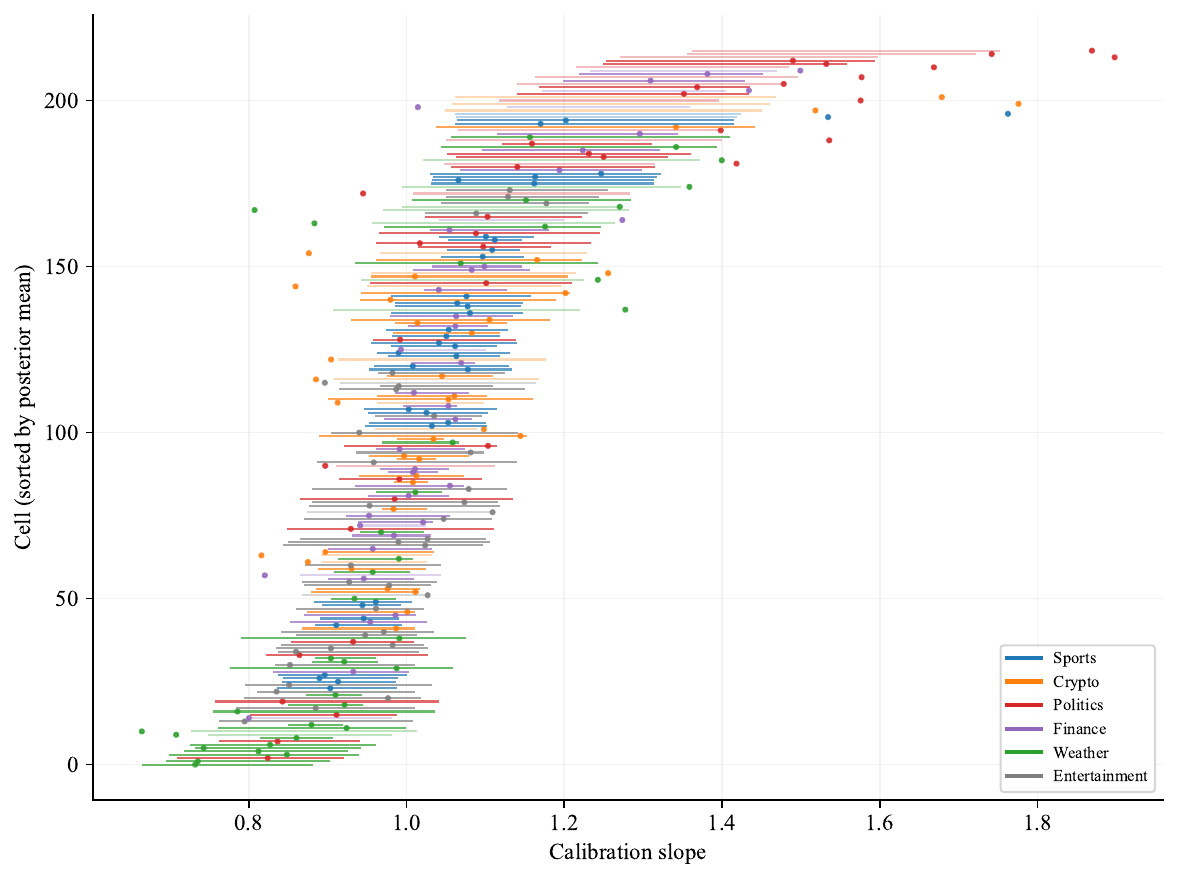}
\caption{Posterior predictive check: observed versus posterior-predicted calibration slopes with 95\% intervals. Of 216 cells, 215 (99.5\%) fall within their prediction intervals.}\label{fig:ppc}
\end{figure}

Of 216 cells, 215 (99.5\%) have observed slopes within their 95\% posterior predictive intervals (Table~\ref{tab:ppc_domain} reports per-domain breakdowns). Coverage slightly exceeds the nominal 95\% because the event-clustered measurement variances are, if anything, conservative; the single cell outside is not part of any systematic pattern (Figure~\ref{fig:ppc}).

%% ============================================================
%% SECTION 7: DISCUSSION
%% ============================================================
\section{Discussion}\label{sec:discussion}

\subsection{Why do domains differ?}\label{sec:why}

The finding that domain-by-horizon interactions constitute the largest calibration component (26.0\%) demands explanation. \citet{ottaviani2008favorite} catalog at least seven families of explanation for the favorite--longshot bias: risk-loving preferences, misperception of small probabilities (prospect-theory weighting), heterogeneous beliefs, informed-trader/adverse-selection effects, market-power and limits to arbitrage, transaction costs and fees, and---specific to parimutuel pools---the simultaneous-betting ``cancellation'' mechanism. It bears emphasis at the outset that the cancellation mechanism does \emph{not} apply here: Kalshi and Polymarket are continuous double-auction order-book markets, not parimutuel pools, so explanations resting on pooled odds are inapplicable. The remaining families are not mutually exclusive, and the descriptive evidence here cannot identify a single one; this section therefore offers three interpretive hypotheses tied to the specific domain, horizon and size patterns documented above, while noting which broader family each invokes.

\emph{Polarized disagreement and heterogeneous beliefs.} Political markets attract traders with strong and often opposing convictions. One plausible interpretation, in the heterogeneous-beliefs family, is that aggressive trading on both sides keeps prices compressed relative to eventual outcome frequencies---related to \citeauthor{manski2006interpreting}'s (\citeyear{manski2006interpreting}) result that prices can diverge from mean beliefs under heterogeneous preferences. The evidence narrows but does not pin down the mechanism: political compression is robust across both exchanges and across event-level clustering, the scale amplification is robust only on Kalshi, and---importantly---Polymarket political trades are not smaller than Kalshi's (median 43.5 vs 45 contracts), so a pure ``cannot-split-large-bets'' story does not by itself explain the cross-platform difference. Institutional frictions that differ across the two venues (fees, tick sizes, order-book depth, pseudonymous wallets, settlement) remain candidate moderators (Tables~\ref{tab:cross_scale}--\ref{tab:cross_bootstrap}; Section~\ref{sec:institutions}).

\emph{The signal over-reaction hypothesis.} Weather markets are uniquely overconfident at short horizons, the only domain where prices are too extreme. This likely reflects over-reaction to meteorological signals. When a forecast predicts a storm tomorrow, traders push prices too far: they overshoot what climatological base rates would justify. Weather's low base rate (24.0\% in Table~\ref{tab:kalshi_summary}, reflecting threshold exceedance contracts) means that over-reacting to ``yes'' signals is particularly costly. At longer horizons, where meteorological signals are weaker and base rates dominate, weather markets converge to the universal underconfidence pattern.

\emph{The information convergence hypothesis.} Sports markets are well calibrated at short-to-medium horizons because information is continuous, quantifiable and publicly observable: game scores update in real time, player statistics are tracked in detail, injury reports follow regular schedules. This generates smooth convergence to truth with minimal disagreement, so that prices reach appropriate extremes. At long horizons (1 month+), sports forecasting becomes more speculative, information advantage declines, and the same favorite--longshot underconfidence that affects all domains takes hold (slope 1.74). The sports trajectory illustrates a general principle: calibration quality tracks the richness of the available information environment \citep[cf.][]{sunstein2006infotopia}.
\subsection{Institutional differences between Kalshi and Polymarket}\label{sec:institutions}

Kalshi and Polymarket differ along dimensions that can affect calibration. Kalshi is CFTC-regulated, uses named accounts, imposes position and compliance constraints, and charges fees that vary with contract economics. Polymarket is blockchain-based, globally accessible, pseudonymous and has a different direct fee environment, although traders may still face spreads, gas-related costs, funding costs and liquidity constraints. These differences matter for interpretation. Kalshi's trading fee follows the published schedule $0.07\,C\,p(1-p)$ (rounded up), so it is largest for contracts priced near 50 cents. Applying this formula to each execution, the mean fee is small and similar across domains---roughly 1.2--1.3 cents per contract (Politics 1.18, Sports 1.33)---so fees do not track the calibration differences documented here, consistent with Reviewer~1's observation that no uniform fee pattern is apparent. This is a \emph{price-based fee-exposure measure} computed from the fee formula, not realized fee payments, which are not observed at the trade level (it also does not capture maker/taker treatment or Polymarket's gas and spread costs). Two facts sharpen the comparison. First, political trade sizes are comparable on the two platforms (median 43.5 vs 45 contracts; 94\% of Kalshi political volume comes from large trades), so the stronger Kalshi scale effect is not simply a matter of larger Kalshi bets. Second, on Kalshi only 16\% of large political executions sweep more than one price level within a timestamp, which suggests that multi-level ``walk-the-book'' executions are unlikely to account for most of the effect; this diagnostic cannot, however, rule out same-price child executions of a single parent order (\ref{app:split_orders} reports a burst-window sensitivity check that does). The present data cannot decompose these institutional channels separately, so platform comparisons are used for robustness and interpretation, not causal identification.
\subsection{Practical implications for prediction market consumers}\label{sec:practical}

Prediction markets are increasingly cited by journalists, embedded in news coverage and referenced by policymakers. The findings suggest domain-specific caution. For any (domain, horizon, trade-size) combination, the estimated intercept $\hat{a}$ and slope $\hat{b}$ transform a raw market price $p \in (0,1)$ into a recalibrated probability.
\begin{equation}\label{eq:recalibrate}
p^* = \sigma\!\bigl(\hat{a} + \hat{b} \cdot \mathrm{logit}(p)\bigr),
\end{equation}
where $\sigma(\cdot)$ is the logistic function. When $\hat{a}$ is close to zero and $\hat{b} > 1$, the recalibrated probability is more extreme than the raw price; when $\hat{b} < 1$, it is more moderate. When $\hat{a}$ is not close to zero, the intercept shifts the entire curve and must be retained. The full 216-cell calibration matrix, including intercepts and slopes, is provided as supplementary material.

\emph{For political markets.} A price of 70 cents in a political market one week before resolution should not automatically be reported as 70\%. If the relevant intercept is near zero, applying the domain- and horizon-specific slope $\hat{b} \approx 1.83$ yields the slope-only illustration
\begin{equation}\label{eq:example}
p^* = \frac{0.70^{1.83}}{0.70^{1.83} + 0.30^{1.83}} \approx 0.83.
\end{equation}
The full recalibration uses Equation~\eqref{eq:recalibrate} with the cell-level intercept. Political prediction market prices are often compressed toward 50\%, with favorites underpriced and longshots overpriced, so users should avoid treating raw prices as context-free probabilities.

\emph{For sports markets.} At horizons under one week, sports market prices are reasonably trustworthy (slopes 0.90--1.10). Beyond one month, the favorite--longshot bias appears strongly (slope 1.74).

\emph{For weather markets.} At short horizons, weather prices are if anything too extreme, over-reacting to signals. Consumers of short-horizon weather contracts should recognize that the price may overstate the probability of the predicted outcome. At moderate horizons, weather markets are among the best calibrated.
\subsection{Implications for market design}\label{sec:design}

\emph{Position limits and fee schedules in political markets.} The scale effect suggests that large Kalshi political trades are associated with amplified price compression. Position limits, progressive fees or enhanced liquidity incentives could in principle reduce the influence of concentrated trading, but the present evidence is not sufficient to recommend a particular intervention. Any design change would need to account for the possibility that large trades also contain information.

\emph{Equal-weighted aggregation as a diagnostic.} The finding that trade-weighted calibration is substantially better than contract-weighted calibration in Politics (mean gap 0.33) suggests that equal-weight or capped-weight aggregators may be useful diagnostics in politically polarized domains. Prediction polls \citep{atanasov2017distilling,pennock2001real} provide one comparison class, but direct performance comparisons require matched questions and time horizons.

\emph{Domain-specific credibility indicators.} Exchanges could display calibration track records alongside prices, such as a traffic-light system indicating whether a given domain--horizon combination has historically been well calibrated, compressed, or over-reactive.
\subsection{Implications for forecasting research}\label{sec:research}

These results challenge domain-agnostic calibration studies. The decomposition shows that pooling across domains obscures the dominant source of calibration variation. Future studies should stratify by domain and time horizon. The five-term descriptive framework extends naturally to election models, expert platforms and machine learning probability outputs.
\subsection{Limitations}\label{sec:limitations}

Several limitations should be noted. First, although core findings are compared on two exchanges, both are US-centric platforms; prediction markets in different regulatory environments (e.g.\ Betfair in the UK; see \citealt{smith2006market}) may exhibit different patterns. The Polymarket comparison is limited to three of six domains and affected by timestamp noise at short horizons.

Second, the domain classification is coarse. Within Politics, subcategories have different dynamics (Section~\ref{sec:artefacts}). A finer-grained decomposition would require a more complex model.

Third, trades are not independent observations: multiple trades occur within the same contract and event, and the yes and no contracts of an event are mechanically linked. This is addressed by reporting event-clustered standard errors for the first-stage slopes (roughly 50 times the naive Fisher errors), market- and event-clustered bootstraps for the scale effect, and a measurement-error model (Section~\ref{sec:bayesian}) that propagates the clustered first-stage uncertainty into the decomposition. The headline conclusions survive this more conservative treatment, but their apparent precision is correspondingly reduced. On Kalshi, \texttt{event\_ticker} provides a clean event identifier; the Polymarket event grouping is coarser, so its clustered intervals should be read as indicative.

Fourth, only executions are observed, not trader identities or parent orders. Exact-timestamp burst aggregation reduces but does not eliminate order-splitting concerns. Wallet fragmentation on Polymarket creates an analogous limitation for interpreting position size.

Fifth, logistic recalibration is one approach to measuring calibration, and the slope is only part of it. The revision adds nonparametric checks---binned reliability and isotonic curves, expected calibration error and the Murphy decomposition of the Brier score---which reproduce the slope-based domain ranking (\ref{app:flexible}), and a parallel decomposition of the intercept together with an integrated calibration error that uses both parameters (\ref{app:intercepts}). These agree with the slope-based conclusions, but the main decomposition remains slope-based for parsimony and interpretability \citep{murphy1977reliability,gneiting2007strictly,brier1950verification}.

Sixth, whether these patterns are stable over time is an open question. The decomposition is designed to be re-estimated as new data becomes available.
%% ============================================================
%% SECTION 8: CONCLUSION
%% ============================================================
\section{Conclusion}\label{sec:conclusion}

Prediction markets are useful information aggregation mechanisms, but their calibration is conditional. Using data from Kalshi and Polymarket, this paper shows that calibration varies systematically across event domains, time-to-resolution and trade size. A five-term descriptive decomposition accounts for 87.3\% of Kalshi cell-level slope variation in sample and 71.5\% in a leave-one-cell-out check.

The most stable result is political underconfidence: political prices are often compressed toward 50\%, and this domain-level pattern is also visible on Polymarket. The trade-size result is more institution-specific. Large Kalshi political trades are associated with additional compression, and this association survives market-clustered and exact-timestamp burst checks, but the corresponding Polymarket estimate is smaller and sensitive to the unified data snapshot.

For users of prediction markets, the implication is practical but limited: raw prices should not be treated as context-free probabilities. Domain, horizon and trading concentration provide useful information about how much recalibration may be needed.
\section*{Data availability statement}

\begin{sloppypar}
The Kalshi and Polymarket data used in this study were collected using the data collection framework and pre-collected dataset of \citet{becker2026microstructure}, available at \url{https://github.com/Jon-Becker/prediction-market-analysis/}. The raw data are also publicly accessible through the Kalshi API (\url{https://trading-api.readme.io/}) and the Gamma API/Polygon blockchain indexer. The analysis code, domain classification rules, reviewer-revision diagnostics and the full 216-cell Kalshi calibration matrix are available at \url{https://github.com/namanhzz/prediction-market-calibration} and as supplementary material. All results are locked to a single deposited unified Polymarket snapshot, and every table and figure regenerates from it with one command. The snapshot comprises \texttt{polymarket.parquet} (6{,}157{,}856 bytes; SHA-256 beginning \texttt{890090\dots c645}) and \texttt{polymarket\_ctf.parquet} (2{,}173{,}978{,}796 bytes; SHA-256 beginning \texttt{cb1dbf\dots df75}); the full 64-character checksums are listed in the replication package. The Kalshi results reproduce exactly from the public raw data, and the Polymarket and cross-platform results reproduce exactly from the deposited snapshot. This snapshot was regenerated from the public source and differs from the one underlying the original submission (it carries a larger long-tail of bespoke markets, so the full-dataset Polymarket totals are larger); all Polymarket and cross-platform numbers reported here are tied to this single deposited version, and the substantive conclusions are unchanged.
\end{sloppypar}

\section*{Supplementary material}

Supplementary material is available online at the journal website. It includes (a) the full 216-cell calibration matrix as a CSV file, (b) domain classification rules (560+ pattern rules) and (c) cross-platform comparison tables.

\section*{Acknowledgements}

The author thanks Jonathan Becker for developing the prediction market data collection framework and pre-collected dataset, which greatly facilitated data acquisition for this study.

\section*{Funding}

This research received no specific grant from any funding agency in the public, commercial or not-for-profit sectors.

\section*{Conflict of interest}

The author declares no conflict of interest. The author has no financial interest in Kalshi, Polymarket or any prediction market exchange.
%% ============================================================
%% REFERENCES
%% ============================================================
\bibliographystyle{plainnat}
\bibliography{references}
%% ============================================================
%% APPENDICES
%% ============================================================
\appendix

\section{Robustness checks}\label{app:robustness}

\subsection{First-stage intercepts and cell sizes}\label{app:intercepts}

Table~\ref{tab:intercepts} summarizes the intercept estimates from the 216 first-stage logistic recalibrations. Intercepts are not uniformly negligible, which is why the main text interprets slopes as compression statistics rather than complete calibration summaries. Politics has a mean intercept close to zero, so its slope-based underconfidence interpretation is less affected by directional yes/no bias than some other domains; Crypto carries the largest directional bias ($\overline{|a|} = 0.41$).

To check that the slope-only focus does not distort the substantive conclusions, two additional steps are taken. First, the same five-term Type~I decomposition is fitted to the intercept $a(d,\tau,s)$ in place of the slope. Directional bias is also strongly structured: the domain-by-horizon term explains 62.8\% of intercept variance and the domain main effect 15.6\%, mirroring the slope decomposition. Second, a joint intercept-and-slope \emph{integrated calibration error}, $\mathrm{ICE} = \sum_p f(p)\,|\sigma(\hat a + \hat b\,\mathrm{logit}\,p) - p|$, is computed from the fitted two-parameter recalibration curve and the contract-weighted price distribution $f(p)$ of each cell. This measure is not model-free---it still rests on the logistic curve---but, unlike the slope alone, it reflects both the intercept and the slope. Ranking domains by volume-weighted ICE identifies Politics as by far the most miscalibrated domain (ICE $0.064$), matching the slope-based conclusion; the remaining domains have low ICE and are hard to separate (Sports lowest at $0.013$). Equal-weighting cells instead promotes Crypto, because a few thin cells carry large intercepts, but this reflects low-volume cells rather than the bulk of trading. The full cell-level intercepts, the intercept decomposition and the ICE table are in the replication output.

\begin{table}[!t]
\caption{Summary of first-stage logistic recalibration intercepts by domain.\label{tab:intercepts}}
\centering
\begin{tabular}{lrrrr}
\toprule
Domain & Mean & Median & Mean $|a|$ & Max $|a|$ \\
\midrule
Crypto        & 0.255  & 0.005  & 0.406 & 1.667 \\
Entertainment & -0.156 & -0.187 & 0.186 & 0.487 \\
Finance       & -0.098 & -0.051 & 0.204 & 0.635 \\
Politics      & -0.006 & -0.057 & 0.208 & 0.495 \\
Sports        & 0.043  & -0.033 & 0.106 & 1.017 \\
Weather       & -0.146 & -0.083 & 0.238 & 1.132 \\
\bottomrule
\end{tabular}
\end{table}

All 216 cells satisfy the 200-trade threshold. The smallest cell counts by domain are 472 in Weather, 1,536 in Politics, 2,049 in Entertainment, 2,678 in Crypto, 5,163 in Finance and 22,518 in Sports.

\subsection{Flexible calibration checks}\label{app:flexible}

The nonparametric validation summarized in the main text (Section~\ref{sec:nonparam}, Table~\ref{tab:nonparam}, Figure~\ref{fig:reliability}) is computed here in full. In addition to the by-domain estimators, binned reliability and isotonic curves are computed by domain~$\times$~coarse-horizon (under-24h, 1d--1w, over-1w). As a representative point, at a raw price of 0.75 the by-domain isotonic estimate is 0.886 for Politics, 0.776 for Sports, 0.746 for Crypto, 0.772 for Finance, 0.691 for Weather and 0.738 for Entertainment, so political compression and short-horizon weather overconfidence are not artifacts of imposing a linear logit slope. The full reliability, isotonic and metric tables are in the replication output.

\subsection{Potential split orders}\label{app:split_orders}

Kalshi trade records identify executions but not parent orders. To assess whether large orders split across order-book levels drive the trade-size result, trades are re-aggregated into exact-timestamp bursts: same contract, timestamp, taker side and execution price. The trade-size bin is then assigned using the summed burst size. Under this conservative aggregation, the Politics large-minus-single effect falls from 0.531 to 0.417 but remains positive with a 95\% bootstrap confidence interval $[0.141, 0.664]$; the Sports effect remains close to zero at 0.082 $[-0.058, 0.268]$. Because same-price child executions could also be split across nearby (not identical) timestamps, the burst window is widened as a sensitivity check: aggregating executions of the same contract and side that share a price within a one-second and a five-second window gives a Politics gap of $+0.42$ and $+0.45$ respectively (exact-timestamp $+0.41$). The effect is therefore stable and positive across increasingly aggressive de-fragmentation; this does not rule out all order-splitting, but it shows the Politics result is not an artifact of trade fragmentation at the timestamp resolution observed.

\subsection{Contract availability over horizons}\label{app:availability}

Some domains contain short-lived contracts that are not available at all horizons. As a robustness check, domain-by-time slopes are re-estimated after restricting to contracts with at least 30 days between open and close. Political slopes are stable on average (mean difference $+0.068$ relative to baseline). Sports slopes become moderately more compressed (mean difference $+0.275$), while Crypto and Weather change substantially because the restriction leaves a small and atypical subset of long-duration contracts in those domains.

A stricter balanced-panel version requested in review is also run: restricting to contracts actually \emph{traded} in at least six of the nine horizon bins, so each contributing contract spans most of the horizon range. Requiring trades in \emph{all nine} fine bins is infeasible---it would discard nearly every short-lived contract (hourly crypto markets, same-day sports lines) and leave too few trades per cell to estimate a slope. To answer the reviewer's request directly, the horizon is therefore coarsened to four bins ($<$24h, 1d--1w, 1w--1mo, $>$1mo) and an \emph{exactly balanced} panel is constructed of the 2{,}859 contracts traded in \emph{all four} bins. On this fully balanced panel, political underconfidence is present at every horizon (slopes 1.34, 1.24, 1.34, 1.63) and the sports pattern is preserved (0.89 rising to 1.37), whereas Crypto and Weather are erratic on the small, atypical long-lived subset that survives the restriction---the same message as the six-of-nine and 30-day checks. The six-of-nine panel is reported as the finer-grained operational approximation to ``available across the horizon range.'' Political slopes are again the most stable (mean change $-0.075$, mean absolute change 0.217), and Sports is stable in level (mean absolute change 0.117), whereas Crypto ($-0.280$) and Weather ($-0.369$) shift substantially, confirming that their horizon profiles are tied to contract design. The main text therefore treats horizon comparisons in Crypto and Weather as conditional on the available contract design rather than as statements about contracts that are uniformly available at every horizon, while the political-underconfidence and sports patterns are robust to the balanced-panel restriction.

\begin{table}[!t]
\caption{Variance decomposition under alternative specifications. This table preserves the original four-block grouping, with common and domain-specific size effects grouped together for comparability across robustness runs.\label{tab:robust}}
\centering
\resizebox{\columnwidth}{!}{
\begin{tabular}{lccccc}
\toprule
Specification & $\mu$ $R^2$ & $\alpha$ $R^2$ & $\beta$ $R^2$ & $\gamma$ $R^2$ & Total $R^2$ \\
\midrule
Baseline $[5,95]$, $C=10$ & 0.302 & 0.146 & 0.260 & 0.165 & 0.873 \\
Price $[2,98]$            & 0.308 & 0.157 & 0.231 & 0.190 & 0.885 \\
Price $[1,99]$            & 0.279 & 0.157 & 0.242 & 0.183 & 0.861 \\
Price $[10,90]$           & 0.295 & 0.097 & 0.333 & 0.137 & 0.861 \\
$C = 1$                   & 0.302 & 0.146 & 0.260 & 0.165 & 0.873 \\
$C = 100$                 & 0.302 & 0.146 & 0.260 & 0.165 & 0.873 \\
\bottomrule
\end{tabular}
}

\smallskip\noindent Total $R^2$ is stable across all price ranges (0.861--0.885) and regularization strengths (identical to three decimals at $C = 1, 10, 100$). The wider price range $[2, 98]$ slightly increases total $R^2$ to 0.885; the domain ranking is preserved.
\end{table}

\begin{table}[!t]
\caption{Scale effect: trade-level, market-clustered and event-clustered bootstrap. Clustering by event (\texttt{event\_ticker}) groups the related contracts of one event and the yes/no sides.\label{tab:bootstrap}}
\centering
\begin{tabular}{llccc}
\toprule
Domain & Bootstrap method & Mean $\Delta$ & 95\% CI & Sig.?\ \\
\midrule
Politics & Trade-level       & $+0.53$ & $[0.29, 0.75]$    & Yes \\
Politics & Market-clustered  & $+0.59$ & $[0.14, 1.32]$    & Yes \\
Politics & Event-clustered   & $+0.63$ & $[0.12, 1.80]$    & Yes \\
Sports   & Trade-level       & $+0.07$ & $[-0.07, 0.26]$   & No  \\
Sports   & Market-clustered  & $+0.01$ & $[-0.03, 0.05]$   & No  \\
Sports   & Event-clustered   & $+0.01$ & $[-0.03, 0.06]$   & No  \\
\bottomrule
\end{tabular}

\smallskip\noindent The Politics effect excludes zero under all three resampling schemes; the event-clustered interval is widest (only 850 event clusters). These event-clustered standard errors are propagated into the measurement-error model (Section~\ref{sec:bayesian}).
\end{table}

\begin{table}[!t]
\caption{Size $\times$ horizon confounding diagnostic: median hours to close by domain and trade size.\label{tab:confound}}
\centering
\begin{tabular}{lcccc}
\toprule
Domain & Single & Small & Medium & Large \\
\midrule
Sports        & 2.3   & 7.6    & 1.5    & 96.3   \\
Crypto        & 202.2 & 5.7    & 0.6    & 5,936.7\\
Politics      & 862   & 1,505  & 1,488  & 213    \\
Finance       & 1.7   & 427.1  & 0.5    & 4.5    \\
Weather       & 14.9  & 15.8   & 24.2   & 156.6  \\
Entertainment & 144.3 & 0.5    & 0.6    & 55.5   \\
\bottomrule
\end{tabular}

\smallskip\noindent In Politics, large trades have shorter median horizons (213 hours) than single-contract trades (862 hours), meaning any horizon confounding would bias against finding a scale effect.
\end{table}
\begin{table}[!t]
\caption{$F$-test derivation and effect sizes.\label{tab:ftest}}
\centering
\begin{tabular}{lccccr}
\toprule
Source & SS & df & MS & $F$ & Partial $\eta^2$ \\
\midrule
$\alpha$ (domain)          & 1.435 & 5   & 0.287 & 33.16 & 0.535 \\
$\kappa$ (size)            & 0.318 & 3   & 0.106 & 12.24 & 0.203 \\
$\beta$ (domain$\times$time) & 2.562 & 40  & 0.064 & 7.40  & 0.673 \\
$\gamma$ (domain$\times$size)& 1.306 & 15  & 0.087 & 10.05 & 0.512 \\
Residual                   & 1.247 & 144 & 0.009 & ---   & ---   \\
\bottomrule
\end{tabular}

\smallskip\noindent The size main effect has $p < 10^{-6}$; the other displayed tests have $p < 10^{-15}$. Partial $\eta^2 = \mathrm{SS}_{\text{component}} / (\mathrm{SS}_{\text{component}} + \mathrm{SS}_{\text{residual}})$.
\end{table}
\section{Bayesian model specification and diagnostics}\label{app:bayesian}

\begin{table}[!t]
\caption{Prior specification.\label{tab:priors}}
\centering
\begin{tabular}{llcl}
\toprule
Parameter & Prior & Dim.\ & Constraint \\
\midrule
$\mu(\tau)$  & $\mathcal{N}(1.0, 0.5)$    & 9  & None \\
$\sigma_\alpha$ & $\mathrm{HalfCauchy}(1.0)$ & 1  & $> 0$ \\
$\alpha_{\mathrm{raw}}$ & $\mathcal{N}(0, 1)$ & 5  & Sum-to-zero \\
$\sigma_\beta$  & $\mathrm{HalfCauchy}(1.0)$ & 1  & $> 0$ \\
$\beta_{\mathrm{raw}}$  & $\mathcal{N}(0, 1)$ & 40 & Doubly centered \\
$\sigma_\delta$ & $\mathrm{HalfCauchy}(1.0)$ & 1  & $> 0$ \\
$\delta_{\mathrm{raw}}$ & $\mathcal{N}(0, 1)$ & 6  & None \\
$\sigma$        & $\mathrm{HalfCauchy}(1.0)$ & 1  & $> 0$ \\
\bottomrule
\end{tabular}

\smallskip\noindent The model uses non-centered parameterizations for all hierarchical effects. The domain intercepts satisfy a sum-to-zero constraint. The interaction matrix $\beta$ is doubly centered.
\end{table}

\begin{table}[!t]
\caption{Hyperparameter posterior summaries (measurement-error model).\label{tab:hyper}}
\centering
\begin{tabular}{lcccl}
\toprule
Parameter & Mean & SD & 95\% CrI & Interpretation \\
\midrule
$\sigma_\alpha$ & 0.093 & 0.048 & $[0.041, 0.216]$ & Domain spread \\
$\sigma_\beta$  & 0.085 & 0.015 & $[0.058, 0.117]$ & Well-constrained \\
$\sigma_\delta$ & 0.033 & 0.017 & $[0.014, 0.074]$ & Small \\
$\sigma$        & 0.006 & 0.004 & $[0.000, 0.017]$ & Residual structural $\approx 0$ \\
\bottomrule
\end{tabular}

\smallskip\noindent CrI denotes a 95\% credible interval. $\sigma$ is the residual structural standard deviation \emph{beyond} the first-stage measurement variance $\mathrm{se}_{d\tau s}^2$ that enters Equation~\eqref{eq:bayesian} directly; its near-zero posterior means little non-measurement variation is detected under this specification. Maximum $\hat{R}$ across all parameters is 1.0000; minimum bulk effective sample size is 3{,}794; no divergent transitions.
\end{table}

\begin{sidewaystable}
\caption{Posterior $\beta$ matrix (domain $\times$ time interaction).\label{tab:beta}}
\centering
\begin{tabular}{lccccccccc}
\toprule
Domain & 0--1h & 1--3h & 3--6h & 6--12h & 12--24h & 24--48h & 2d--1w & 1w--1mo & 1mo+ \\
\midrule
Sports        & $+.089$ & $-.042$ & $-.061$ & $+.018$ & $+.049$ & $+.029$ & $-.062$ & $-.033$ & $+.012$ \\
Crypto        & $+.014$ & $+.051$ & $+.085$ & $-.039$ & $-.033$ & $+.065$ & $-.057$ & $-.123$ & $+.036$ \\
Politics      & $-.051$ & $-.111$ & $-.067$ & $+.067$ & $-.017$ & $+.020$ & $+.052$ & $+.071$ & $+.036$ \\
Finance       & $+.001$ & $+.071$ & $+.036$ & $-.026$ & $-.031$ & $-.116$ & $-.022$ & $+.105$ & $-.017$ \\
Weather       & $-.019$ & $-.046$ & $-.082$ & $-.031$ & $+.017$ & $+.038$ & $+.092$ & $-.026$ & $+.057$ \\
Entertainment & $-.034$ & $+.076$ & $+.089$ & $+.011$ & $+.015$ & $-.036$ & $-.002$ & $+.006$ & $-.125$ \\
\bottomrule
\end{tabular}

\smallskip\noindent The $\beta$ matrix captures each domain's slope deviation from the universal horizon mean $\mu(\tau)$ plus the domain intercept $\alpha$.
\end{sidewaystable}

\begin{table}[!t]
\caption{Posterior predictive coverage by domain.\label{tab:ppc_domain}}
\centering
\begin{tabular}{lccc}
\toprule
Domain & Cells & Within 95\% & Coverage \\
\midrule
Sports        & 36 & 36 & 100.0\% \\
Crypto        & 36 & 36 & 100.0\% \\
Politics      & 36 & 35 & 97.2\% \\
Finance       & 36 & 36 & 100.0\% \\
Weather       & 36 & 36 & 100.0\% \\
Entertainment & 36 & 36 & 100.0\% \\
\midrule
\textbf{All}  & \textbf{216} & \textbf{215} & \textbf{99.5\%} \\
\bottomrule
\end{tabular}

\smallskip\noindent Coverage exceeds the nominal 95\% because the event-clustered first-stage measurement variances entering Equation~\eqref{eq:bayesian} are conservative.
\end{table}
\section{Cross-platform validation details}\label{app:crossplatform}

This appendix provides full tabular results for the cross-platform comparison between Kalshi and Polymarket. Polymarket trade timestamps are derived from Polygon block numbers via a bucketed lookup table with approximately 6-hour granularity, introducing approximately 3 hours of noise. Bins 0--1 hour and 1--3 hours are therefore unreliable for cross-platform comparison and are excluded from mean slope calculations.

\begin{table*}[!t]
\caption{Cross-platform calibration slopes ($\Delta$ = Polymarket $-$ Kalshi). Bins marked $\dagger$ are unreliable due to timestamp noise.\label{tab:cross_slopes}}
\centering
\resizebox{\columnwidth}{!}{
\begin{tabular}{lccclccclccc}
\toprule
& \multicolumn{3}{c}{Sports} && \multicolumn{3}{c}{Crypto} && \multicolumn{3}{c}{Politics} \\
\cline{2-4}\cline{6-8}\cline{10-12}
Bin & Kalshi & Poly & $\Delta$ && Kalshi & Poly & $\Delta$ && Kalshi & Poly & $\Delta$ \\
\midrule
0--1h$\dagger$  & 1.101 & 1.026 & $-0.076$ && 0.993 & 1.059 & $+0.066$ && 1.341 & 0.699 & $-0.642$ \\
1--3h$\dagger$  & 0.960 & 1.187 & $+0.227$ && 1.013 & 0.875 & $-0.138$ && 0.933 & 1.255 & $+0.322$ \\
3--6h           & 0.897 & 0.917 & $+0.020$ && 1.065 & 0.973 & $-0.092$ && 1.317 & 1.514 & $+0.197$ \\
6--12h          & 1.006 & 1.093 & $+0.087$ && 1.007 & 0.991 & $-0.016$ && 1.552 & 1.216 & $-0.336$ \\
12--24h         & 1.053 & 0.996 & $-0.057$ && 1.006 & 1.114 & $+0.108$ && 1.477 & 1.392 & $-0.085$ \\
24--48h         & 1.075 & 1.090 & $+0.015$ && 1.209 & 1.091 & $-0.118$ && 1.515 & 1.204 & $-0.311$ \\
2d--1w          & 1.037 & 1.027 & $-0.010$ && 1.121 & 0.990 & $-0.131$ && 1.833 & 2.077 & $+0.244$ \\
1w--1mo         & 1.240 & 0.952 & $-0.288$ && 1.090 & 1.162 & $+0.072$ && 1.833 & 1.671 & $-0.162$ \\
1mo+            & 1.740 & 1.336 & $-0.404$ && 1.357 & 1.072 & $-0.285$ && 1.730 & 1.083 & $-0.647$ \\
\midrule
Mean (rel.)     & 1.150 & 1.059 & $-0.091$ && 1.122 & 1.056 & $-0.066$ && 1.608 & 1.451 & $-0.157$ \\
\bottomrule
\end{tabular}
}

\smallskip\noindent\emph{Note:} Slopes regenerated from the locked Polymarket snapshot (Section~\ref{sec:bayesian}; Data Availability). Mean (rel.)\ is the unweighted mean slope across the seven reliable time bins (3--6h through 1mo+). Political underconfidence replicates on Polymarket (mean 1.45); the two shortest bins ($\dagger$) are unreliable due to block-number timestamp noise.
\end{table*}

\begin{table}[!t]
\caption{Cross-platform scale effect by domain and trade size.\label{tab:cross_scale}}
\centering
\resizebox{\columnwidth}{!}{
\begin{tabular}{llccccc}
\toprule
Domain & Platform & Single & Small & Medium & Large & $\Delta$(L$-$S) \\
\midrule
Sports   & Kalshi     & 1.002 & 1.010 & 1.007 & 1.013 & $+0.011$ \\
Sports   & Polymarket & 1.141 & 1.043 & 1.076 & 1.048 & $-0.093$ \\
Crypto   & Kalshi     & 1.028 & 1.025 & 1.023 & 1.004 & $-0.024$ \\
Crypto   & Polymarket & 1.007 & 1.065 & 1.058 & 1.069 & $+0.062$ \\
Politics & Kalshi     & 1.188 & 1.224 & 1.373 & 1.741 & $+0.554$ \\
Politics & Polymarket & 1.211 & 1.051 & 1.407 & 1.393 & $+0.182$ \\
\bottomrule
\end{tabular}
}

\smallskip\noindent\emph{Note:} $\Delta$(L$-$S) is the difference between aggregate Large and Single slopes (pooled across all time bins). This differs from the bootstrap estimand in Equation~\eqref{eq:delta}, which averages within-horizon differences: Kalshi Politics aggregate $\Delta = +0.554$ vs within-horizon $\Delta = +0.531$; Polymarket Politics aggregate $+0.182$ vs within-horizon $+0.281$. The Kalshi gradient is monotonic in trade size; the Polymarket gradient is not.
\end{table}

\begin{table}[!t]
\caption{Cross-platform whale effect bootstrap.\label{tab:cross_bootstrap}}
\centering
\begin{tabular}{llccc}
\toprule
Platform \& method & Domain & $\Delta$(L$-$S) & 95\% CI & Sig.?\ \\
\midrule
Kalshi, cell             & Politics & $+0.531$ & $[+0.288, +0.747]$ & Yes \\
Kalshi, market-clustered & Politics & $+0.593$ & $[+0.144, +1.318]$ & Yes \\
Kalshi, event-clustered  & Politics & $+0.630$ & $[+0.115, +1.795]$ & Yes \\
Polymarket, cell             & Politics & $+0.281$ & $[+0.026, +0.542]$ & Yes \\
Polymarket, market-clustered & Politics & $+0.210$ & $[-0.311, +1.122]$ & No  \\
Polymarket, market-clustered & Sports   & $-0.100$ & $[-0.284, +0.083]$ & No  \\
Polymarket, market-clustered & Crypto   & $+0.060$ & $[-0.021, +0.138]$ & No  \\
\bottomrule
\end{tabular}

\smallskip\noindent\emph{Note:} The Kalshi Politics scale effect is significant under cell, market- and event-clustered resampling. The Polymarket Politics effect is positive and significant at the cell level but \emph{not} once trades are clustered by market, so the trade-size amplification is specific to Kalshi.
\end{table}
The Polymarket Politics 2-day-to-1-week bin (slope 2.077) is a local maximum, plausibly reflecting a concentration of 2024 US election daily-resolution markets whose peak trading fell 2--7 days before resolution. Several limitations merit note: the regex-based domain classifier assigns a large share of Polymarket markets to ``Other''; Weather and Entertainment are effectively absent on Polymarket; the count field represents contracts per on-chain transaction rather than full position size; and the comparison is locked to the deposited snapshot (Data Availability), which differs from the originally submitted snapshot---the trade and contract counts in the three analyzed domains are essentially unchanged (Politics 45.7M trades in both), while the current snapshot carries a larger long-tail of bespoke ``Other'' markets, so full-dataset Polymarket totals are larger---and yields the same qualitative conclusions.
\section{Additional supplementary appendices}\label{app:additional}

The following materials are provided in the supplementary data. (a) Domain classification rules for both Kalshi (560+ ticker-prefix patterns) and Polymarket (compiled regular expression patterns on market titles); (b) the full 216-cell calibration matrix as a CSV file, with observed slopes, standard errors, fitted decomposition components and residuals; and (c) political subcategory analysis, including slope ranges for 10 subcategories and a Simpson's paradox diagnostic at the 1--3 hour horizon.

\end{document}